\documentclass[aps,prd,onecolumn,groupedaddress,showpacs,nofootinbib,amssymb]{revtex4}
\usepackage[dvips]{graphicx}
\usepackage{amssymb}
\usepackage{amsmath}
\usepackage{graphicx}
\usepackage{amsfonts}
\usepackage{bm}

\begin{document}

\title{A quantitative analysis of singular inflation with scalar-tensor and modified gravity}
\author{
S.~Nojiri,$^{1,2}$\,\thanks{nojiri@gravity.phys.nagoya-u.ac.jp}
S.~D.~Odintsov,$^{3,4,6}$\,\thanks{odintsov@ieec.uab.es}
V.~K.~Oikonomou,$^{5,6}$\,\thanks{v.k.oikonomou1979@gmail.com}}
\affiliation{
$^{1)}$ Department of Physics, Nagoya University, Nagoya 464-8602, Japan \\
$^{2)}$ Kobayashi-Maskawa Institute for the Origin of Particles and the
Universe, Nagoya University, Nagoya 464-8602, Japan \\
$^{3)}$Institut de Ciencies de lEspai (IEEC-CSIC),
Campus UAB, Carrer de Can Magrans, s/n\\
08193 Cerdanyola del Valles, Barcelona, Spain\\
$^{4)}$ ICREA,
Passeig LluA­s Companys, 23,
08010 Barcelona, Spain\\
$^{5)}$ Department of Theoretical Physics, Aristotle University of Thessaloniki,
54124 Thessaloniki, Greece\\
$^{6)}$ National Research Tomsk State University, 634050 Tomsk,
Tomsk State Pedagogical University, 634061 Tomsk, Russia}

\begin{abstract}
We provide a detailed quantitative description of singular inflation. Its close analogy
with finite-time future singularity which is associated to dark energy era is described. Calling and classifying the singularities of such
inflation as  finite-time cosmological singularities we investigate their occurrence, with special emphasis on the Type IV singularity.
The study is performed in the context of a general non-canonical scalar-tensor theory. In addition, the impact of finite time singularities on the
slow-roll parameters is also investigated. Particularly, we study three cases, in which the singularity occurs during the inflationary era, at the end,
and also we study the case that the singularity occurs much more later than inflation ends. Using the obtained slow-roll parameters, for each case,
we calculate explicitly the spectral index of primordial curvature perturbations $n_s$, the associated running of the spectral index $a_s$ and of the scalar-to-tensor ratio $r$ and compare the resulting values to the Planck and BICEP2 data. As we demonstrate, in some cases corresponding to the Type IV singularity, there might be the possibility of agreement with the observational data, when the singularity occurs at the end, or after inflation. However, absolute concordance of all observational indices is not achieved. On the contrary, if the singularity occurs during the inflationary era, this is catastrophic for the theory, since the observational indices become divergent. We also show how a Type IV singularity may be consistently accommodated in the Universe's late time evolution, and we study the behavior of the effective equation of state. In addition, we investigate when inflation ends classically, in the context of our scalar-tensor model. Finally, we investigate which $F(R)$ gravity can generate the Type IV singularity, with special emphasis on the behavior near the finite time singularity.
\end{abstract}

\pacs{04.50.Kd, 95.36.+x, 98.80.-k, 98.80.Cq,11.25.-w}

\maketitle



\def\pp{{\, \mid \hskip -1.5mm =}}
\def\cL{\mathcal{L}}
\def\be{\begin{equation}}
\def\ee{\end{equation}}
\def\bea{\begin{eqnarray}}
\def\eea{\end{eqnarray}}
\def\tr{\mathrm{tr}\, }
\def\nn{\nonumber \\}
\def\e{\mathrm{e}}

\section*{Introduction}

The appearance of singularities in classical cosmology is an inevitable inherent feature that almost all the theoretical descriptions carry along. In the Einstein-Hilbert description of local gravity effects, singularities are hidden behind the horizons of black holes, so the appearance of cosmological singularities at a large scale is rendered a rather unwanted feature of classical cosmological theories. Therefore, one of the most important quests in current cosmological research is to fully understand the presence of singularities, either these appear in the far past or in the far future. In the former case, the singularity is usually called the initial singularity, and a very fundamental question is posed, that is, whether the Universe really started in a singular state and continued its evolution entering an inflationary era \cite{inflation}, or whether the Universe is described by some sort of a bounce \cite{bounce}, in which case there is no initial singularity. The initial singularity is one point in spacetime at which spacetime is geodesically incomplete. This type of singularities is perfectly described by the Hawking-Penrose singularity theorems, given some time ago \cite{hawkingpenrose}. The geodesic incompleteness is practically ensured by the existence of time-like and null geodesics that cannot be extended to arbitrary values of their parameters. This constraint is equivalent to the general and stringent requirements of the Hawking-Penrose theorems, which are epitomized by the strong energy condition $R^{\mu \nu}K_{\mu}K_{\nu}\geq 0$, for every non-space-like vector $K_{\mu}$, and also from the non existence of closed time-like curves. In addition, on these spacetime singularities, some of the curvature invariants diverge. However, sudden and Type II singularities that occur at a finite future time, can be quite milder from the initial spacetime singularity, in the sense that these are points at which it is possible that some observable quantities (for example the Hubble rate or its higher derivatives) might become unbounded at some finite time, but this does not necessarily implies geodesic incompleteness. This type of singularities was studied by Barrow \cite{barrowsing1} some time ago, and further developed in \cite{barrowsing2,barrowsing3,barrow,Barrow:2015ora}. According to Barrow's description of the sudden singularities, at these singularities the strong energy and weak energy condition are satisfied, contraction is not required and the scale factor, along with its first derivative are finite. However, the second derivative of the scale factor and the pressure blow up at the finite time which these singularities correspond to. In addition, there exist sudden singularities of generalized type, at which the singular behavior occurs for higher derivatives of the scale factor. In principle, although that sudden singularities are scalar polynomial curvature singularities, these are much weaker than the Big Rip singularities, since geodesics are extendible through these points. In addition, these are not singularities of the crushing type, such as the Big Rip singularity \cite{Nojiri:2005sx,Caldwell:2003vq}.

Another consistent and important classification of future cosmological singularities, in which the value of observable quantities is taken into account, was given in \cite{Nojiri:2005sx}. These singularities occur frequently in the context of modified gravity theories \cite{reviews1}. According to this classification, the observable quantities that are taken into account are the scale factor, the effective energy density, the effective pressure, the Hubble rate and its higher derivatives. It is obvious that some singularities of this type may not necessarily lead to geodesic incompleteness, thus can, in principle, be milder in some sense from the Universe's initial singularity. The initial singularity and perhaps the Big Rip singularity are crushing singularities, in the following sense: at some time during the evolution, each point that lies on the spacelike three dimensional hypersurface $t= \mathrm{const}$ corresponds to a unique geodesic curve. Loosely speaking, these geodesic world lines form a congruence of geodesics that if these diverge in the past, or converge in the future to a unique point, then this point is essentially a singularity where the geodesic incompleteness occurs.

Recently in Ref.~\cite{Barrow:2015ora}, Barrow and Graham studied a large family of finite-time cosmological singularities for a canonical scalar field with a power-law potential. Using a thorough qualitative analysis, they demonstrated that these weak finite-time singularities are an inherent feature of cosmological models with similar matter content. In addition, for appropriate initial conditions these models can describe large field inflation, with the difference that the physical system evolves to a singular state when inflation ends. Motivated by this work, we provide a detailed quantitative description of the singular inflation analysis, and classify the singularities found in \cite{Barrow:2015ora}, according to the cosmological singularities classification given in \cite{barrowsing1,barrowsing2,barrowsing3,Barrow:2015ora}. As we shall demonstrate, there is some difference in our analysis, in comparison to the one presented in \cite{Barrow:2015ora}, traced in the case at which higher derivatives of the Hubble rate are taken into account. This feature characterizes the Type IV singularities, according to the finite time cosmological singularities classification given in \cite{Nojiri:2005sx}. More importantly, as we explicitly show, the existence of a Type IV singularity in an inflationary model can have dramatic effects on the slow-roll inflation parameters, rendering them singular at finite time, if the singularity occurs during inflation. We have to note that it is possible the slow-roll inflation does not work near the Type IV singularity. Indeed, the corresponding approximation may be broken especially, if the singularity occurs during inflation. Moreover, if inflation ends exactly at a Type IV singularity, this can be compatible with current observation data, and interestingly enough, after the Type IV singularity, the Universe is accelerated in a quintessential way. In addition, we study the evolution of the effective equation of state as a function of cosmic time and we also address the question when inflation ends, providing a detailed quantitative analysis. With regards to the question when inflation ends, we found a particularly interesting solution which occurs in the case of a Type IV singularity, according to the finite time cosmological singularities classification provided in \cite{Nojiri:2005sx}. As we show, there appears the physically appealing picture that, if certain requirements are met, inflation ends, classically at least, at a non-singular state at a time $t_e$, before the Type IV singularity occurs, and the Type IV singularity is reached in a future time, $t_s$, with $t_s\gg t_e$. Thus in a Universe that evolves in time with a Hubble rate that has a Type IV singularity, inflation can end in a non-singular state.

In the study of future Universe after $\Lambda$CDM-like epoch, the finite-time future singularities that may occur, are classified according to Ref.~\cite{Nojiri:2005sx}. However, most attention was given to Type I (Big Rip) future singularity (for incomplete list of references on that, see Refs.~\cite{Caldwell:2003vq,ref5}) or Type II singularity, commonly known as a sudden future singularity (for an incomplete list of references on that, see Ref.~\cite{barrow}). These future singularities may occur for phantom, or quintessence-like, dark energy Universe, even if currently it shows nearly $\Lambda$CDM-like evolution. As the classification of these singularities is applied to accelerating dark energy Universe, it is evident that it may be applied in the same sense to the classification of singularities after singular inflation. Hence, in this paper we adopt just the same notations for all four types of singularities, even after inflation. Of course, one should bear in mind that it is not physically the same as future singularities after $\Lambda$CDM epoch, because finite-time singularity may occur prior to dark energy epoch or after current dark energy epoch. In addition, it may also occur after dark energy epoch. In this respect, we shall focus on the Type IV future singularity, because it is expected that Universe may survive after passing through such singularity smoothly, since it is not a crushing type of singularity. Thereby, it could be possible that between (singular perhaps) inflation and  the $\Lambda$CDM epoch, the Universe passed through type IV singularity. Furthermore, for unified inflation-dark energy models of the sort introduced in Ref.~\cite{Nojiri:2005pu}, we may get the following really physical appealing possibility. The future singularity of Type II, III or IV, may occur just because of the epoch of singular inflation, so that dark energy era could be de Sitter-like or pure $\Lambda$CDM. In other words, the future singularity may be predicted by only the study of unified inflation-dark energy cosmology, like in the examples of Ref.~\cite{Nojiri:2005pu}. However, by studying only the dark energy epoch, one discovers that it is $\Lambda$CDM-like and therefore, the future universe is regular. Thereby, based only on the analysis of dark energy epoch, one may arrive to wrong conclusions about the future universe. We shall exemplify a scenario like this in a section, later on in this paper.

Finally, owing to the fact that the finite time cosmological singularities classified according to Refs.~\cite{Nojiri:2005sx}, frequently occur in the context of modified gravity, we investigate which $F(R)$ gravity can generate the Type IV singularity, providing the exact form of the $F(R)$ gravity near the Type IV singularity. As is expected, the resulting $F(R)$ gravity is approximately of polynomial type near the singularity, which can be fine-tuned to be of the Einstein-Hilbert form gravity plus curvature corrections.

This paper is organized as follows: In section I, we quantify Barrow's singular inflation description, in the context of scalar-tensor gravity, by using very well known scalar-tensor reconstruction techniques \cite{Nojiri:2005pu,Capozziello:2005tf}. Using very general arguments, we thoroughly describe all the singularity types that can occur in these theories, and we exemplify our results using illustrative examples. In addition, we investigate the effects of a singularity in the slow-roll inflation parameters and also study the behavior of the effective equation of state parameter as a function of the cosmic time. Moreover, the question when inflation classically ends is addressed too. In section II, we investigate a physically appealing scenario, according to which, in the end of inflation, the Universe experiences a Type IV singularity and in addition the dark energy era is quintessential or approximately de Sitter and at finite time after the dark energy era, the Universe experiences again a Type IV singularity. In section III we investigate which $F(R)$ gravity can realize the Type IV singularity, which proves to be the most relevant type of singularities. We address this issue using qualitative arguments and we quantify our study by using very well known reconstruction techniques, in order to find the $F(R)$ gravity near the Type IV singularity. In section IV we briefly investigate the impact of the finite time singularities on the long-wavelength scalar perturbation modes, with special emphasis in the Type IV singularity. In section V we discuss the graceful exit problem and this is affected by the presence of finite time singularities. Our conclusions along with a brief critical discussion, follow in the end of paper.

\section{Scalar-tensor gravity analysis}

In this section we shall provide a quantitative analysis of the singular inflation physics and its consequences on observable data, in the context of a general scalar-tensor gravity theory containing one scalar field. For related studies, the reader is referred to Refs.~\cite{Nojiri:2005sx,sergnoj}.

We start off our analysis by recalling how the finite time cosmological singularities are classified according to Refs.~\cite{Nojiri:2005sx}. For a detailed presentation of these issues consult also \cite{reviews}. As was shown in Ref.~\cite{Nojiri:2005sx,sergnoj}, the finite-time future singularities are classified as follows,
\begin{itemize}
\item Type I (``Big Rip'') : When $t \to t_s$, the scale factor diverges $a$,
the effective energy density $\rho_\mathrm{eff}$, the effective pressure
$p_\mathrm{eff}$ diverge, $a \to \infty$, $\rho_\mathrm{eff} \to \infty$, and
$\left|p_\mathrm{eff}\right| \to \infty$.
This type of singularity was presented in Ref.~\cite{Caldwell:2003vq} and studied in Ref.~\cite{Nojiri:2005sx}.
\item Type II (``sudden'') \cite{barrow}: When $t \to t_s$, the scale factor and the effective energy density is finite, $a \to a_s$, $\rho_\mathrm{eff} \to \rho_s$ but
the effective pressure diverges $\left|p_\mathrm{eff}\right| \to \infty$.
\item Type III : When $t \to t_s$, the scale factor is finite, $a \to a_s$ but the effective
energy density and the effective pressure diverge, $\rho_\mathrm{eff} \to \infty$,
$\left|p_\mathrm{eff}\right| \to \infty$.
\item Type IV : For $t \to t_s$, the scale factor, the effective energy density, and
the effective pressure are finite, that is, $a \to a_s$, $\rho_\mathrm{eff} \to \rho_s$,
$\left|p_\mathrm{eff}\right| \to p_s$, but the higher derivatives of the Hubble rate
$H\equiv \dot a/a$ diverge.
\end{itemize}
The Type III and Type IV singularities were  extensively studied in Ref.~\cite{Nojiri:2005sx}. Notice that, the effective energy density $\rho_\mathrm{eff} $ and the effective pressure
$p_\mathrm{eff}$ are defined in the following way,
\be
\label{IV}
\rho_\mathrm{eff} \equiv \frac{3}{\kappa^2} H^2 \, , \quad
p_\mathrm{eff} \equiv - \frac{1}{\kappa^2} \left( 2\dot H + 3 H^2
\right)\, .
\ee
In Ref.~\cite{Barrow:2015ora}, a scalar field model realizing inflation with a Type IV singularity has been
proposed. In this section, by using the general formulation of the scalar reconstruction of Refs.~\cite{Nojiri:2005pu,Capozziello:2005tf}, we construct general
scalar-tensor models that realize Type II and IV singularities. We consider the following non-canonical scalar-tensor action, which describes a general single scalar field model,
\be
\label{ma7}
S=\int d^4 x \sqrt{-g}\left\{
\frac{1}{2\kappa^2}R - \frac{1}{2}\omega(\phi)\partial_\mu \phi
\partial^\mu\phi - V(\phi) + L_\mathrm{matter} \right\}\, .
\ee
The function $\omega(\phi)$ is called the kinetic function, $V(\phi)$ is the scalar potential, and both are assumed to be functions of the scalar field $\phi$. In addition, we assume a flat Friedmann Robertson Walker (FRW) metric of the form,
\begin{equation}
\label{metricformfrwhjkh}
\mathrm{d}s^2=-\mathrm{d}t^2+a^2(t)\sum_i\mathrm{d}x_i^2\, .
\end{equation}
The function $\omega(\phi)$ in Eq.~(\ref{ma7}) is irrelevant and we can therefore absorb this function by redefining the scalar field $\phi$, in the following way,
\be
\label{ma13}
\varphi \equiv \int^\phi d\phi \sqrt{\omega(\phi)} \,,
\ee
where we assumed that $\omega(\phi)>0$. Then the kinetic term of the scalar field in the action (\ref{ma7}) is rewritten as,
\be
\label{ma13b}
 - \omega(\phi) \partial_\mu \phi \partial^\mu\phi
= - \partial_\mu \varphi \partial^\mu\varphi\, .
\ee
It is convenient for later purposes, if we keep $\omega(\phi)$, although we
can absorb $\omega(\phi)$ into the redefinition of the scalar field. In the case that  $\omega(\phi)<0$, the scalar field becomes ghost corresponding to phantom dark energy and in this case, instead of (\ref{ma13}), by using the redefinition,
\be
\label{ma13p}
\varphi \equiv \int^\phi d\phi \sqrt{-\omega(\phi)} \, ,
\ee
we find that,
\be
\label{ma13bp}
 - \omega(\phi) \partial_\mu \phi \partial^\mu\phi
= \partial_\mu \varphi \partial^\mu\varphi\, ,
\ee
instead of the expression given in Eq.~(\ref{ma13b}). In the case that the scalar field is a ghost field, the energy density becomes unbounded from below in the classical theory, but in the quantum theory, the energy becomes always bounded from below, with the cost that there appears a negative norm.

 From the action of Eq.~(\ref{ma7}), we obtain the following expressions of the
energy density and the pressure : \be \label{ma8} \rho =
\frac{1}{2}\omega(\phi){\dot \phi}^2 + V(\phi)\, ,\quad p =
\frac{1}{2}\omega(\phi){\dot \phi}^2 - V(\phi)\, . \ee Therefore,
the scalar potential $V(\phi)$ and the parameter $\omega(\phi)$ can
be written in terms of the Hubble rate and its derivatives as
follows, \be \label{ma9} \omega(\phi) {\dot \phi}^2 = -
\frac{2}{\kappa^2}\dot H\, ,\quad
V(\phi)=\frac{1}{\kappa^2}\left(3H^2 + \dot H\right)\, . \ee

Consider the following solution,
 \be \label{ma11} \phi=t\, ,\quad
H=f(t)\, . \ee If we assume that $\omega(\phi)$ and $V(\phi)$ are
given by using the single function $f(\phi)$, as follows, \be
\label{ma10} \omega(\phi)=- \frac{2}{\kappa^2}f'(\phi)\, ,\quad
V(\phi)=\frac{1}{\kappa^2}\left(3f(\phi)^2 + f'(\phi)\right)\, , \ee
we find that the solution (\ref{ma11}) is indeed a solution of the
FRW equations of the system.

It can be easily verified that the equation which results by the
variation of the action, with respect to $\phi$, namely, \be
\label{ma12} 0=\omega(\phi)\ddot \phi +
\frac{1}{2}\omega'(\phi){\dot\phi}^2 + 3H\omega(\phi)\dot\phi +
V'(\phi)\, , \ee is also satisfied by the solution (\ref{ma11}). In
effect, the arbitrary Universe's evolution, expressed by
$H(t)=f(t)$, can be realized by an appropriate choice of
$\omega(\phi)$ and $V(\phi)$. In other words, by defining the
particular type of the Universe's evolution, the corresponding
scalar-Einstein gravity may be found.

In the case of the Type II and IV singularities,
the Hubble rate $H(t)$ may be chosen in the following form:
\be
\label{IV1}
H(t) = f_1(t) + f_2(t) \left( t_s - t \right)^\alpha\, .
\ee
Here $f_1(t)$ and $f_2(t)$ are smooth (differentiable) functions of $t$ and $\alpha$
is a constant. If $0<\alpha<1$, there appears Type II singularity and if $\alpha$ is
larger than $1$ and not integer, there appears Type IV singularity.
When $\alpha$ is given by two integers $n$ and $m$ by
\be
\label{IV2}
\alpha= \frac{n}{2m + 1}\, ,
\ee
we can consider the region where $t>t_s$.

Then, by using (\ref{ma10}), we find
\begin{align}
\label{IV3}
\omega(\phi) = & - \frac{2}{\kappa^2}
\left\{ f_1'(\phi) + f_2'(\phi) \left( t_s - \phi \right)^\alpha
 - \alpha f_2( \phi ) \left( t_s - \phi \right)^{\alpha - 1} \right\}\, , \nn
V(\phi) = & \frac{1}{\kappa^2}\left\{
3\left(  f_1( \phi ) + f_2( \phi ) \left( t_s - \phi \right)^\alpha \right)^2
+ f_1'(\phi) + f_2'(\phi) \left( t_s - \phi \right)^\alpha
 - \alpha f_2(\phi) \left( t_s - \phi \right)^{\alpha - 1} \right\}\, ,
\end{align}
which determine a general model which generates the Type IV singularity when
$\alpha>1$.  When $t=\phi \sim t_s$, if $\alpha>1$, the expression of $\omega(\phi)$ in (\ref{IV3}) gives
\be
\label{Ph1}
\omega(\phi) \sim - \frac{2}{\kappa^2} f_1'(t_s)\, .
\ee
Hence, although the scalar field is canonical if $f'_1(t_s)<0$, the scalar field becomes ghost, corresponding to the phantom dark energy if $f'_1(t_s)>0$.
In other words, the Type IV singularity is realized not only in the case that the scalar field is canonical, but also the singularity can be realized even in the phantom phase, where the phantom dark energy dominates.

We first consider the simple case that $f_1(t)=0$ and $f_2(t)=f_0$ with a positive constant
$f_0$. In the neighborhood of $t=t_s$, we find that,
\be
\label{IV4}
\omega(\phi) = \frac{2\alpha f_0}{\kappa^2}
\left( t_s - \phi \right)^{\alpha - 1} \, , \quad
V(\phi) \sim - \frac{\alpha f_0 }{\kappa^2}
\left( t_s - \phi \right)^{\alpha - 1} \, ,
\ee
and by using (\ref{ma13}), we find
\be
\label{IV5}
\varphi = - \frac{2 \sqrt{2\alpha f_0}}{\kappa \left(\alpha+ 1 \right) }
\left( t_s - \phi \right)^{\frac{\alpha + 1}{2}}\, ,
\ee
Consequently, the scalar potential reads,
\be
\label{IV6}
V(\varphi) \sim - \frac{\alpha f_0 }{\kappa^2}
\left\{ - \frac{\kappa \left(\alpha+ 1 \right) }{2 \sqrt{2\alpha f_0}} \varphi
\right\}^{\frac{2 \left( \alpha - 1 \right)}{\alpha + 1}}\, .
\ee
Therefore, when the following condition holds true,
\be
\label{IV6b}
 -2 < \frac{2 \left( \alpha - 1 \right)}{\alpha + 1}<0\, ,
\ee
there occurs the Type II singularity. Accordingly, the Type IV singularity occurs when the following holds true,
\be
\label{IV7}
0< \frac{2 \left( \alpha - 1 \right)}{\alpha + 1}<2\, .
\ee
The above results seem to be a little bit different from those appearing in
\cite{Barrow:2015ora}. As was qualitatively shown in \cite{Barrow:2015ora}, the potential is of the form
$V(\varphi) \sim \varphi^n$
and therefore we may identify,
\begin{equation}
\label{estraident}
n= \frac{2 \left( \alpha - 1 \right)}{\alpha + 1} \, .
\end{equation}
In the claim of \cite{Barrow:2015ora}, when $1>n>0$, the Type IV singularity could occur, in which case $\ddot H$ diverges, and when
$k+1>n>k$, the $k$-th derivative of $H$, that is $H^{(k+2)}$, diverges.
Then the Type IV singularity should occur, even if $n>2$, which is in conflict with Eq.~(\ref{IV7}).
In the qualitative analysis of \cite{Barrow:2015ora}, it has been assumed that $\dot\varphi\neq 0$ at $t=t_s$ when $3>\alpha>2$, that is,
$\ddot H$ is finite, but $H^{(3)}$ diverges at $t=t_s$, but as we can find from (\ref{IV5}), $\dot\varphi$ behaves as $\dot\varphi \sim \left( t_s - t \right)^{\frac{\alpha - 1}{2}}$ and therefore $\dot \phi$ vanishes at $t=t_s$.
This shows that the arguments in \cite{Barrow:2015ora} might not apply in our case, and specifically when $3>\alpha>2$ or equivalently $n>2$. In addition, the Type IV singularity does not
occur when $n>2$.

We should note that in  Eq.~(\ref{IV1}),
\begin{itemize}
\item $\alpha<-1$ corresponds to the Type I singularity.
\item $-1<\alpha<0$ corresponds to Type III singularity.
\item $0<\alpha<1$ corresponds to Type II singularity.
\item $\alpha>1$ corresponds to Type IV singularity.
\end{itemize}
In the case that $f_1(t)=0$ and $f_2(t)=f_0$, we have assumed that $f_0$ should
be a positive constant in order for the Hubble rate $H$ given in Eq.~(\ref{IV1}) to be
positive. Then in the case that $\alpha$ is negative, which corresponds to the Type I and II
singularities, $\omega(\phi)$ in Eq.~(\ref{IV4}) and therefore the scalar field becomes ghost corresponding to a phantom era.
In the case that $\alpha$ is positive, which corresponds to the Type II and IV
singularities, the potential $V(\varphi)$ in (\ref{IV4}) becomes unbounded from below,
but the expression in (\ref{IV4}) is only valid at $t\sim t_s$ and the full
expression in (\ref{IV3}) gives,
\be
\label{IVA1}
V(\phi) = \frac{1}{\kappa^2}\left\{ 3f_0^2 \left( t_s - t \right)^{2\alpha}
 - \alpha f_0 \left( t_s - t \right)^{\alpha - 1} \right\}\, ,
\ee
which is bounded from below if $\alpha>1$.
Even if $\alpha<1$, by choosing $f_1(\phi)$ and $f_2(\phi)$ properly, we can
make $V(\varphi)$ bounded from below and even to be positive semi-definite.
In the case that $f_1(t)=0$ and $f_2(t)=f_0$, the expression
$n= \frac{2 \left( \alpha - 1 \right)}{\alpha + 1}$ implies that,
\begin{itemize}
\item $n>2$ corresponds to the Type I singularity.
\item $-2<n<1$ corresponds to Type III singularity.
\item $n<-2$ corresponds to Type II singularity.
\item $1<n<2$ corresponds to Type IV singularity.
\end{itemize}

Even in case that $f_1(t)=0$ and $f_2(t)=f_0$, with $f_0$ a positive constant, in Eq.~(\ref{IV4}), the slow-roll inflation could occur, and at the end of
the inflationary era there could occur the Type IV singularity. We shall elaborate on this further later on in this section.

We may consider more general models, by choosing a more
general form for the arbitrary functions $f_1(t)$ and $f_2(t)$. In this way, we can realize almost all kind of the cosmologies possessing a Type IV singularity at $t=t_s$.
For example, if we choose,
\be
\label{IV8}
f_1(t)= \frac{f_1}{\sqrt{t_0^2 + t^2}}\, , \quad
f_2(t) = \frac{f_2t^2}{t_0^4 + t^4}\, ,
\ee
or more generally,
\be
\label{IV8b}
f_1(t)= \frac{f_1}{\left(t_0^n + t^n\right)^{\frac{1}{n}}}\, , \quad
f_2(t) = \frac{f_2t^2}{t_0^4 + t^4}\, ,
\ee
the Universe becomes a de Sitter spacetime, with the Hubble rate $H$ becoming a constant, that is, $H\sim \frac{f_1}{t_0}$ in the early Universe. In addition, $H$ behaves as $H\sim \frac{f_1}{t}$ at late time, although the Type IV singularity appears at $t=t_s$.
If we choose $f_2$ small enough, the second term (\ref{IV1}) can always be neglected, only if we consider the development in the expansion of the Universe. This argument is quite general, because we can always choose the second term to be small enough, compared to the first term in (\ref{IV1}). Therefore, the Type IV singularity is not always relevant to the expansion history of the Universe. However, as we will now explicitly demonstrate, the Type IV singularity affects strongly the slow-roll parameters, if the singularity occurs during the inflationary era.

The slow-roll parameters $\epsilon$, $\eta$ and $\xi$ by $H$ can be
expressed as follows \cite{inflation,inflationreview,nojserghybrid},
\begin{align}
\label{S7}
\epsilon =& \frac{1}{2\kappa^2} \left(\frac{d\phi}{d\varphi} \right)^2
\left( \frac{V'(\phi)}{V(\phi)} \right)^2
= \frac{1}{2\kappa^2} \frac{1}{\omega(\phi)}
\left( \frac{V'(\phi)}{V(\phi)} \right)^2 \, , \nn
\eta = & \frac{1}{\kappa^2 V(\phi)} \left[ \frac{d\phi}{d\varphi}
\frac{d}{d\phi}
\left( \frac{d\phi}{d\varphi} \right) V'(\phi)
+ \left( \frac{d\phi}{d\varphi} \right)^2 V''(\phi) \right]
=  \frac{1}{\kappa^2 V(\phi)} \left[ - \frac{\omega'(\phi)}{2 \omega(\phi)^2}
V'(\phi) + \frac{1}{\omega(\phi)} V''(\phi) \right] \, , \nn
\xi^2 = & \frac{V'(\phi)}{\kappa^4 V(\phi)^2 \omega(\phi)^2 }
\left\{ \left[ - \frac{\omega''(\phi)}{2\omega(\phi)}
+ \left( \frac{\omega'(\phi)}{\omega(\phi)} \right)^2 \right]V' (\phi)
 - \frac{3\omega'(\phi)}{2\omega(\phi)} V''(\phi) + V'''(\phi)
\right\} \, .
\end{align}
We should note that the derivatives of $V(\phi)$ and $\omega(\phi)$ diverge at $\phi=t=t_s$ in general.

There could be three cases,
\begin{enumerate}
\item\label{cI} The Type IV singularity occurs during the inflationary era.
\item\label{cII} The inflationary era ends with the Type IV singularity.
\item\label{cIII} The Type IV singularity occurs after the inflationary era.
\end{enumerate}

We first consider the case \ref{cI}, that is, the case in which the singularity occurs during inflation.
Then, the above slow-roll parameters may diverge, which may be in conflict with the current observational data. In the same line of research, there are several cases, in which the aforementioned argument holds true, for example,
\begin{itemize}
\item When $f_1(t_s)=f_1'(t_s)=0$, $V(\phi)$ and $\omega(\phi)$ behave as $\left(t_s - t \right)^{\alpha-1}$ near the singularity as in (\ref{IV4}).
Therefore we find,  $\epsilon$, $\eta \sim \left(t_s - t \right)^{- \alpha-1}$, $\xi^2 \sim \left(t_s - t \right)^{- \alpha-3}$, and therefore
all the slow-roll parameters diverge at $t=t_s$.
\item When $f_1(t_s)\neq 0$ but $f_1'(t_s)=0$,  $\omega(\phi)$ behaves as $\left(t_s - t \right)^{\alpha-1}$, again but $V(\phi)$ is finite if $\alpha>1$.
Furthermore if $2>\alpha>1$, $V'(\phi) \sim \left(t_s - t \right)^{\alpha-2}$ and therefore $\epsilon \sim \left(t_s - t \right)^{\alpha-3}$, $\eta \sim \left(t_s - t \right)^{- 2}$, $\xi^2 \sim \left(t_s - t \right)^{- 4}$, and therefore all the slow-roll parameters diverge at $t=t_s$.
\end{itemize}
All the parameters $\epsilon$, $\eta$, and $\xi^2$ are finite for $\alpha>4$, and also when,
$f_1(t_s)$, $f_1'(t_s)$, $f_1''(t_s)$, $f_1^{(3)}(t_s)$, and $f_1^{(4)}(t_s)$ do not
vanish. In this case, however, these parameters do not depend on $f_2(t_s)$ and its derivatives, and consequently the slow-roll parameters, do not depend on the singularity.

In all the above considerations, we have assumed that the Type IV singularity may occur during the inflationary era, as in case \ref{cI}.
We may, of course, as in case \ref{cII} or \ref{cIII}, consider models that the singularity could occur after the inflationary era.
If there is any matter coupled with the higher derivative of the curvature, which includes the higher derivatives of $H$, the Type IV singularity may affect the matter content, but owing to the fact that such a coupling is not renormalizable, this could be relevant only at a very high energy region.


Let us now address in detail the remaining cases \ref{cII} or \ref{cIII}. We start off with case \ref{cII}, in which case inflation is assumed to end at the time $t_f$ which is equal to $t_f=t_s$, with $t_s$, which appears in the Hubble rate of Eq.~(\ref{IV1}), being the time at which the possible singularity may occur. We also assume that $f_1(t)=0$ and $f_2(t)=f_0$, so that the Hubble parameter and the corresponding scalar potential near the singularity $t\sim t_s$, are equal to,
\begin{equation}\label{hubblepar}
H(t)=f_0\left( t_s-t\right)^{\alpha}\, , \quad V(\phi)=- \alpha f_0 \left( t_s - \phi \right)^{\alpha - 1}\, .
\end{equation}
It is worth working using the canonical scalar field $\varphi$, instead of $\phi$, in order to make direct contact with the model of Ref.~\cite{Barrow:2015ora}. In terms of the canonical field $\varphi$, the scalar potential near the singularity is equal to,
\begin{equation}\label{varphipotential}
V(\varphi )\simeq V_0\varphi^n\, ,
\end{equation}
with $n$ given in Eq.~(\ref{estraident}), but we quote it again for simplicity,
\begin{equation}\label{neqndef}
n= \frac{2 \left( \alpha - 1 \right)}{\alpha + 1}\, .
\end{equation}
In addition, $V_0$ in Eq.~(\ref{varphipotential}), is equal to,
\begin{equation}\label{vodef}
V_0=-\frac{\alpha f_0}{\kappa^2}\left(-\frac{\kappa (\alpha+1)}{2\sqrt{\alpha}f_0}\right)^{\frac{2 \left( \alpha - 1 \right)}{\alpha + 1}}\, .
\end{equation}
In terms of $n$, the Hubble rate is written as follows,
\begin{equation}\label{hubbleraten}
H(t)=f_0\left( t_s-t\right)^{-\frac{2+n}{-2+n}}\, .
\end{equation}
We shall investigate the impact of the fact that $t_f=t_s$, on the inflation slow-roll parameters and consequently to the observational indices corresponding to inflation. The slow-roll parameters we shall study are the ones given in Eq.~ (\ref{S7}), which when defined in terms of the canonical scalar field $\varphi$ are equal to,
\begin{equation}\label{slowrollind}
\epsilon=\frac{1}{2\kappa^2}\left(\frac{V'(\varphi)}{V(\varphi)}\right)\, , \quad \eta=\frac{1}{\kappa^2}\left(\frac{V''(\varphi)}{V(\varphi)}\right)\, , \quad \xi^2
=\frac{1}{\kappa^2}\left(\frac{V'(\varphi)V'''(\varphi)}{V(\varphi)^2}\right)\, ,
\end{equation}
with the prime this time denoting differentiation with respect to $\varphi$. It is worth to recall at this point how the slow-roll parameters are derived for the case of a canonical scalar field. For a detailed account on these issues, consult Refs.~\cite{inflation,inflationreview}. In general, the slow-roll condition relies upon the constraint,
\begin{equation}\label{slowrollconstr}
\frac{1}{2}\dot{\varphi}^2\ll V(\varphi)\, ,
\end{equation}
and that this constraint holds true for an extended period of time. This condition ensures a long and finite acceleration era and it is known as the first slow-roll condition. In order for the first slow-roll condition to last over an extended period of time, a second condition must imposed,
\begin{equation}\label{slowrollconstrnew}
|\ddot{\varphi}|\ll \left| \frac{\partial V(\varphi)}{\partial \varphi} \right|\, ,
\end{equation}
which is known as the second slow-roll condition. In virtue of the equation of motion satisfied by the canonical scalar field in a flat FRW background, the condition (\ref{slowrollconstrnew}) can be written as,
\begin{equation}\label{newcondqweeee}
|\ddot{\varphi}|\ll 3H |\dot{\varphi}| \, .
\end{equation}
With the two slow-roll conditions holding true, the canonical scalar field equations for the scalar field reads,
\begin{equation}\label{eqnsmotoinscla}
\dot{\varphi }\simeq -\frac{1}{3H}\frac{\partial V(\varphi )}{\partial \varphi} \, ,
\end{equation}
while the FRW equations become,
\begin{equation}\label{frweqnbecomes}
3 H^2\simeq 8\pi G V(\varphi ) \, .
\end{equation}
It can be shown that the first and second slow-roll conditions, namely Eqs.~(\ref{slowrollconstrnew}) and (\ref{newcondqweeee}), are equivalent to the following relations,
\begin{equation}\label{sfinalfgwajd}
\left(\frac{V '(\varphi )}{V (\varphi )}\right)\ll 2\kappa^2\, , \quad
\left(\frac{V'' (\varphi )}{V (\varphi ^2)}\right)\ll \kappa^2 \, ,
\end{equation}
with the prime denoting differentiation with respect to the canonical scalar field. The above equations can be written as follows,
\begin{equation}\label{sloweqnsderivation}
\epsilon \ll 1\, , \quad \eta \ll 1\, ,
\end{equation}
where $\epsilon$ and $\eta$ are the slow-roll parameters,
\begin{equation}\label{slowrolpparaenedefs}
\epsilon =\frac{1}{2\kappa^2}\left(\frac{V '(\varphi )}{V (\varphi )}\right),{\,}{\,}{\,}\eta =\frac{1}{\kappa^2}\left(\frac{V''(\varphi )}{V (\varphi ^2)}\right) \, .
\end{equation}
Notice that the conditions (\ref{sloweqnsderivation}) are actually the slow-roll conditions.

The observational quantities we shall study are the spectral index of the primordial curvature fluctuations $n_s$, the tensor-to-scalar ratio $r$, the associated running of the spectral index $a_s$, which are defined in terms of the slow-roll parameters as follows \cite{inflation,inflationreview},
\begin{equation}\label{observquantit}
n_s\sim 1-6\epsilon+2\eta\, , \quad r=16\epsilon\, , \quad a_s\sim 16 \epsilon \eta-24\epsilon^2-2\xi^2\, .
\end{equation}

The latest Planck observational data \cite{planck} constrain the above observational quantities as follows,
\begin{equation}\label{planckconstr}
n_s=0.9603\pm 0.0073\, , \quad r<0.11\, , \quad a_s=-0.0134\pm 0.009\, ,
\end{equation}
and the BICEP2 \cite{BICEP2} observations strongly indicate that,
\begin{equation}\label{bicep2}
r=0.2^{+0.07}_{-0.05}\, .
\end{equation}
Note that the Planck data are currently considered to be more reliable than the BICEP2 results. Finally, the values of the scalar field at the beginning and at the end of inflation, $\varphi_i$ and $\varphi_f$ correspondingly, are chosen is such a way so that the Lyth bound \cite{lyth} is respected, so that,
\begin{equation}\label{lythb}
|\kappa\varphi_i-\kappa\varphi_f|\geq 1\, ,
\end{equation}
when the BICEP2 results are taken into account, while consistency with the Planck data requires,
\begin{equation}\label{planck}
|\kappa\varphi_i-\kappa\varphi_f|< 1\, .
\end{equation}
Practically, these requirements imply that the number of $e$-folds until inflation ends, is $N\simeq 50$ for the BICEP2 result, while for the Planck, $N\simeq 60$.
Having all the above at hand, we can perform a full quantitative analysis of the observational implications that a Hubble rate and the corresponding scalar potential of the form given in Eq.~(\ref{hubblepar}) would have. A similar detailed analysis for the case of chaotic inflation potentials was performed in \cite{nojserghybrid}. For the potential of Eq.~(\ref{hubblepar}), the slow-roll parameters are equal to,
\begin{equation}\label{slowrollpar}
\epsilon(\varphi)=\frac{n^2}{2 \kappa^2 (t_s-\varphi )^2}\, ,\quad \eta(\varphi)=\frac{(-1+n) n}{\kappa ^2 (t_s-\varphi )^2 \text{}}\, ,\quad \xi^2=\frac{(-2+n) (-1+n) n^2}{\kappa ^4 (t_s-\varphi )^4 \text{}}\, .
\end{equation}

So let us now investigate what will happen if $t_f=t_s$. Under the slow-roll approximation, inflation will actually end, when the scalar field acquires it's final value $\varphi_f$, at which the slow-roll parameter $\epsilon(\varphi_f)$ (or $\eta(\varphi)$)) will be of the order $\epsilon(\varphi_f)\sim 1$. From Eq.~(\ref{slowrollpar}), this would imply for $\varphi_f$ that,
\begin{equation}\label{impl1}
t_s-\varphi_f=\frac{n}{\sqrt{2} \kappa}\, .
\end{equation}
In order to have a clear picture for the ending of inflation, we shall express the slow-roll parameters as a function of the number of $e$-folds until the end of inflation, which is defined to be equal to,
\begin{equation}\label{nefold1}
N=-\int_{t_f}^{t}H(t)\mathrm{d}t\, ,
\end{equation}
and for the Hubble rate given in Eq.~(\ref{hubbleraten}), this is equal to,
\begin{equation}\label{efold2}
N=\frac{1}{4} (-2+n) (-t+t_s)^{-\frac{4}{-2+n}}-\frac{1}{4} (-2+n) (-t_f+t_s)^{-\frac{4}{-2+n}}\, ,
\end{equation}
which is clearly divergent for $t_f=t_s$, if $n>2$. Recall that the $e$-folding number is constructed in such a way so that, when inflation ends, it has to be exactly equal to zero, with it's value increasing as $\varphi$ tends to it's initial value $\varphi_i$. Therefore, we find a first point of inconsistency caused for some values of $n$. Recall from our previous classification of singularities that the $n>2$ case corresponds to the Big Rip singularity, so in the case of a Big Rip singularity, we come across to a severe divergence of the $e$-folding number. So the Big Rip case is an idiomorphic case to be excluded from our study, at least when case \ref{cII} is considered. Therefore, in the following we shall assume that $n<2$, at least in the context of case \ref{cII}.

In the slow-roll approximation, the $e$-folding number $N$, may be approximated by the following relation \cite{inflation,inflationreview,nojserghybrid},
\begin{equation}\label{slowrollnapprox}
N\simeq \kappa^2\int_{\varphi_f}^{\varphi}\frac{V(\hat{\varphi})}{V'(\hat{\varphi})}\mathrm{d}\hat{\varphi}\, ,
\end{equation}
which for the scalar potential of Eq.~(\ref{varphipotential}), reads,
\begin{equation}\label{efoldhubslowroll}
N\simeq \kappa^2 \left(-\frac{t_s \varphi }{n }+\frac{\varphi ^2}{2 n}+\frac{t_s \varphi_f}{n }-\frac{\varphi_f^2}{2 n }\right)\, .
\end{equation}
Substituting the value of $\varphi_f$ from Eq.~(\ref{impl1}) in the above equation, the $e$-folding number becomes,
\begin{equation}\label{infefoldfinal}
N\simeq -\frac{n^2-2 \kappa ^2 (t_s-\varphi )^2}{4 n}\, ,
\end{equation}
so by solving the above equation with respect to $t_s-\varphi$ we obtain,
\begin{equation}\label{fneqnsa}
t_s-\varphi=\sqrt{\frac{4Nn+n^2}{2k^2}}\, .
\end{equation}
We can substitute $t_s-\varphi$ appearing in Eq.~(\ref{fneqnsa}), to the slow-roll parameters of Eq.~(\ref{slowrollpar}), and finally we will have the slow-roll parameters as functions of the $e$-folding number for the whole inflationary era. By doing that we obtain,
\begin{equation}\label{finalslowrollefolds}
\epsilon\simeq \frac{n }{4 N+n}\, ,\quad \eta\simeq \frac{2(-1+n) }{4 N+n }\, ,\quad \xi^2=\frac{8(-2+n ) (-1+n )}{ (4 N+n)^2 }\, ,
\end{equation}
so that finally, the observational indices read,
\begin{equation}\label{finalobservindices}
n_s\simeq \frac{4 (N-1)-n}{4N+n}\, ,\quad r\simeq\frac{16n}{4N+n},{\,}{\,}{\,}a_s\simeq -\frac{8(n+2)}{(4N+n)^2}\, .
\end{equation}
Having these at hand, we can promptly examine which values of $n$, are ruled out by observations, by taking the $e$-folding number to be $N=60$, $N=50$ or $N=54$.
\begin{table}[!]
\begin{center}
\resizebox{\columnwidth}{!}
{
\begin{tabular}{|c|c|c|c|}
\hline
Observational Indices & Spectral Index $n_s$ & Tensor-to-Scalar ratio $r$  & Running of Spectral Index $a_s$
\\\hline
$N=50$, $n=1.9$ & $n_s\simeq 0.961367$ & $r\simeq 0.15057  $ & $a_s\simeq -0.000765389$  \\
\hline
$N=50$, $n=1.382$ & $n_s\simeq 0.966412$ & $r\simeq 0.109801$ & $a_s\simeq -0.000727579 $  \\
\hline
$N=50$, $n=0.5$ & $n_s\simeq 0.975062$ & $r\simeq 0.0399002$ & $a_s\simeq -0.000497509$  \\
\hline
$N=54$, $n=1.48$ & $n_s\simeq 0.967997$ & $r\simeq 0.108884 $ & $a_s\simeq -0.000588614$  \\
\hline
$N=54$, $n=1.8$ & $n_s\simeq 0.965106 $ & $r\simeq 0.132231$ & $a_s\simeq -0.000640852$  \\
\hline
$N=54$, $n=0.9$ & $n_s\simeq 0.97326$ & $r\simeq 0.06639$ & $a_s\simeq -0.000493138$  \\
\hline
$N=60$, $n=1.7$ & $n_s\simeq 0.969384$ & $r\simeq 0.112536$ & $a_s\simeq -0.000506685$  \\
\hline
$N=60$, $n=1.1$ & $n_s\simeq 0.974285$ & $r\simeq 0.0729988$ & $a_s\simeq -0.000426636$  \\
\hline
$N=60$, $n=0.9$ & $n_s\simeq 0.978387 $ & $r\simeq 0.0399002$ & $a_s\simeq -0.000359312$  \\
\hline
 \end{tabular}}
\end{center}
\caption{\label{case2alphamegal}Values of the spectral index of primordial curvature perturbations $n_s$, the associated running of the spectral index $a_s$ and of the scalar-to-tensor ratio $r$, for various values of the $e$-folding number $N$ and of the parameter $n$, for the case that the singularity occurs exactly at the end of inflation (Case \ref{cII}). As it can be seen, the Type IV singularity gives results that are compatible to BICEP2 and Planck data, when the spectral index $n_s$ and the scalar-to-tensor ratio $r$ is considered}
\end{table}
By looking at the Table \ref{case2alphamegal}, it is obvious that, for $(N,n)=(50,1.382)$, $(N,n)=(54,1.48)$ and $(N,n)=(54,1.8)$, the results are compatible with the BICEP2 observational data, with regards to the spectral index $n_s$ and the scalar-to-tensor ratio $r$. However, in all the cases, the running of the spectral index is smaller than the Planck one. For $(N,n)=(50,1.382)$ and $(N,n)=(54,1.48)$, the spectral index $n_s$ and the ratio $r$ are compatible to the Planck observational data. Recall that when $1<n<2$, the physical system develops a Type IV singularity and by looking Table \ref{case2alphamegal}, it is obvious that, if inflation ends at a Type IV singularity, the results can be compatible to the Planck and BICEP2 observational data.

Now we consider case \ref{cIII}, in which case $t_s$ is considered to occur much more later than the inflation ending time $t_f$, that is, $t_s\gg t_f$. Then, contrary to case \ref{cII}, there is no inconsistency in the definition of the $e$-folding number in Eq.~(\ref{efold2}), since it is always finite, and in addition it is equal to zero when inflation ends, as it should be. So practically all the values of $n$ are allowed and all singularities may occur at late times, with no inconsistency. Let us see how the observational indices change in this case. Following the lines of research of our previous analysis, in the slow-roll approximation, inflation ends at, $\epsilon(\varphi_f)\sim 1$, so in this case we have,
\begin{equation}\label{impl1largea}
t_f-\varphi_f=\frac{(-1+n )}{\sqrt{2} \kappa}\, .
\end{equation}
so in the slow-roll approximation, by substituting $\varphi_f$ from Eq.~(\ref{impl1largea}), the $e$-folding number $N$ reads,
\begin{equation}\label{efoldhubslowrolllargea}
N\simeq -\frac{-2 \sqrt{2} \kappa (t_f-t_s) n+n^2+2 \kappa^2 (t_f-\varphi ) (t_f-2 t_s+\varphi )}{4 n}\, ,
\end{equation}
and by solving with respect to $\varphi$, we get,
\begin{align}\label{fneqnsalargea}
& \varphi=\frac{1}{4 \kappa^2}\left(4 \kappa^2 t_s 
+\sqrt{16 \kappa^4 t_s^2-8 \kappa^2 \Big{(}-1-4 Nn-2 \kappa^2 t_f^2+4 \kappa^2 t_f t_s+2 \alpha +2 \sqrt{2} \kappa (t_f-t_s) n -(n+1)^2}\right)\, .
\end{align}
Substituting the above expression in Eq.~(\ref{slowrollpar}), we obtain the slow-roll parameters, with $\epsilon $ being equal to,
\begin{equation}\label{epslargeac}
\epsilon \simeq \frac{n^2}{1+4 N n+(n+1)^2-2 (n+1)  \left(1+\sqrt{2} (t_f-t_s) \kappa \right)+2 (t_f-t_s) \kappa  \left(\sqrt{2}+t_f \kappa -t_s \kappa \right)}\, .
\end{equation}
Accordingly, $\eta$ is equal to,
\begin{equation}\label{etalargea}
\eta \simeq \frac{2 (n-1) n}{\left(1+4 N n+(n-1)^2-2 (n-1)  \left(1+\sqrt{2} (t_f-t_s) \kappa \right)+2 (t_f-t_s) \kappa  \left(\sqrt{2}+t_f \kappa -t_s \kappa \right)\right) }\, ,
\end{equation}
while, $\xi^2$ reads,
\begin{equation}\label{etaxilarge}
\xi^2 \simeq \frac{4 (n-2 ) (n-1 ) n^2}{\left(1+4 N n+(n-1)^2-2 (n-1) \left(1+\sqrt{2} (t_f-t_s) \kappa \right)+2 (t_f-t_s) \kappa \left(\sqrt{2}+t_f \kappa -t_s \kappa \right)\right)^2 }\, .
\end{equation}
Having the slow-roll indices at hand, we can promptly calculate the observational indices, which are,
\begin{align}
\label{finalobservindiceslargea}
n_s \simeq& \frac{4 (n-1 ) n+\left(4 N n-5 n^2-2 \sqrt{2} (t_f-t_s) n \kappa +2 (t_f-t_s)^2 \kappa ^2\right) }{\left( 1+4 N n+(n+1) ^2-2 (n+1)  \left( 1+\sqrt{2} (t_f-t_s) \kappa \right) +2 (t_f-t_s) \kappa  \left(\sqrt{2}+t_f \kappa -t_s \kappa \right)\right) }\, , \nn
r\simeq& \frac{16 n^2}{1+4 N n+(n+1) ^2-2 (n+1)  \left( 1+\sqrt{2} (t_f-t_s) \kappa \right) +2 (t_f-t_s) \kappa  \left( \sqrt{2}+t_f \kappa -t_s \kappa \right)}\, ,\nn
a_s\simeq&  -\frac{8 n^2 }{\left(1+4 N n+(n+1) ^2-2 (n+1)  \left(1+\sqrt{2} (t_f-t_s) \kappa \right) +2 (t_f-t_s) \kappa \left(\sqrt{2}+t_f \kappa -t_s \kappa \right) \right)^2 }\nn &
\times \left( 6-5 (n+1) +(n+1) ^2+\left( 1-8 N n+(n+1) ^2+(n+1)  \left( -2+4 \sqrt{2} (t_f-t_s) \kappa \right) \right. \right. \nn
&\quad \left. \left. - 4 (t_f-t_s) \kappa  \left( \sqrt{2}+t_f \kappa -t_s \kappa \right) \right) \right) \, .
\end{align}
The above expressions are quite complicated, but can be simplified if certain assumptions hold true. It is quite natural to assume that $t_s$ is a quite large number, so that the singularity occurs much more later than the present time. But even if we choose $t_s$ to be equal to the present age of the Universe, $t_s$ is quite large, so let $t_s=6.636\times 10^{51}$GeV$^{-1}$, which is the present age of the Universe. In addition, $\kappa=8\pi G=2.0944\times 10^{-18}$GeV$^{-1}$, with $G$, Newton's gravitational constant. So obviously, the term $2 \kappa^2 (t_f - t_s)^2$ dominates (it is of the order $10^{68}$) in all relations of Eq.~(\ref{finalobservindiceslargea}), so the observational indices may be simplified as follows,
\begin{align}\label{finalobservindiceslargea11}
& n_s \simeq 1 \, ,\nn & r\simeq \frac{256}{2 \kappa^2 (t_f-t_s)^2}\ll 1\, ,
\nn & a_s\simeq \frac{8 n^2}{\left(k^2 (t_f-t_s)^2\right) } \ll 1\, ,
\end{align}
with $a_s>0$. Therefore, the Planck constraints are not satisfied, when the spectral index $n_s$ and the running of the spectral index $a_s$ are taken into account. Now a question rises, what if the time that the singularity occurs, occurred in the past, but after inflation? In that case, it is possible that the observational constraints are satisfied, for example, if $t_s\sim 2 t_f$, that is, the singularity occurs after the end of inflation, for $(N,n)=(55,1.55)$, we get, $n_s\simeq 0.967953$ and $r\simeq 0.108352$, which are compatible with Planck and BICEP2 results.


In conclusion, we can say that the Type IV singularity offers a good phenomenology, since in all the cases that $n$ contributes in the resulting expressions of the observational indices, some of the values with $1<n<2$, result to observational indices compatible with observations.

Coming back to the paradigms for the functions $f_1(t)$ and $f_2(t)$, instead of choosing (\ref{IV8}), we may choose,
\be
\label{IV8c}
f_1(t)= \frac{f_1}{\sqrt{t_0^2 + t^2}} + f_0\, , \quad
f_2(t) = \frac{f_2t^2}{t_0^4 + t^4}\, ,
\ee
where it is assumed that $\frac{f_1}{t_0} \gg f_0 >0$ and also that $f_2$ is small enough. Then, the behavior of the early time is not different from that given in Eq.~(\ref{IV8}) and we find $H\sim \frac{f_1}{t_0}$ but even at late time ($t\to \infty$),
the Hubble rate becomes a constant $H\sim f_0$, which may describe the dark energy in the present Universe. Then the model unifies inflation in the early Universe with the accelerating expansion in the present Universe.
If we choose $t_0$ large enough, the Type IV singularity may occur in the Universe where the dark energy dominates.

We now study the effective equation of state (EoS) parameter $w_\mathrm{eff}$, which is equal to,
\begin{equation}
\label{eos}
w_\mathrm{eff}=\frac{p}{\rho}=-1-\frac{2\dot{H}}{3H^2} \, .
\end{equation}
This parameter will reveal whether the scalar field will go to phantom or not, during the process of cosmological evolution. We shall be mainly interested for the model with Hubble rate given in Eq.~(\ref{IV1}), with $f_1(t)=0$ and $f_2(t)=f_0>0$. For this Hubble parameter, the EoS parameter reads,
\begin{equation}
\label{eos-1}
w_\mathrm{eff}=-1-\frac{2 \alpha(-t+t_s)^{-1-\alpha}}{3f_0} \, .
\end{equation}
Obviously, the EoS parameter can be singular if $\alpha>-1$, and in the converse case it is non-singular. Let us investigate if the EoS evolves to phantom or not. When $\alpha<-1$ and since $f_0>0$, the sign of the fraction appearing in Eq.~(\ref{eos-1}) is determined by the actual value of $-\alpha-1$, for $t>t_s$. We shall consider cosmological times after the singularity, that is, $t>t_s$ and in addition, $\alpha$ is assumed to be of the form given in Eq.~(\ref{IV2}), with $n$ an even integer. Then, when $t>t_s$, and regardless if $\alpha>-1$ or $\alpha<-1$, the EoS describes an quintessence accelerating Universe, since, the EoS is always $w_\mathrm{eff}>-1$. In the case that $\alpha<-1$, the EoS is regular at $t=t_s$, while for $\alpha>-1$, it is always singular. Let us make contact with the singularity study we performed earlier. When $\alpha>-1$, the EoS is always singular at the transition point $t=t_s$. For a cosmological time with $t>t_s$, then the EoS is $w_\mathrm{eff}>-1$, so non-phantom, quintessence acceleration occurs. Notice that when $\alpha>1$, the singularity is Type IV, so the fact that the EoS is always $w_\mathrm{eff}>-1$, has very appealing physical consequences, as we will demonstrate shortly.

In conclusion, the EoS of state analysis revealed a physically appealing case which occurs when $t_s$ is considered to be an early time in the evolution of the Universe. Consequently, a natural question to ask is, is it possible for inflation to end at this Type IV time and more generally, when does inflation ends. As we explicitly demonstrated earlier, if inflation ends exactly at a Type IV singularity, which occurs at $t=t_s$, it may be possible for the theory to be compatible with the observations. Having these issues in mind, we now perform a qualitative analysis of the inflation ending time. A similar analysis on the inflation ending was performed in \cite{sergnoj}. Practically, inflation ends when $\ddot{a}=0$. For the Hubble parameter (\ref{IV1}), the scale factor $a(t)$ reads,
\begin{equation}
\label{scf}
a(t)=\e^{-\frac{f_0 (-t+t_s)^{1+\alpha }}{1+\alpha }}\, ,
\end{equation}
and therefore, $\ddot{a}$ reads,
\begin{equation}
\label{ddscf}
\ddot{a}(t)=\e^{-\frac{f_0 (-t+t_s)^{1+\alpha }}{1+\alpha }} f_0^2 (-t+t_s)^{2 \alpha }-\e^{-\frac{f_0 (-t+t_s)^{1+\alpha }}{1+\alpha }} f_0 (-t+t_s)^{-1+\alpha } \alpha\, ,
\end{equation}
which can be recast as follows,
\begin{equation}
\label{ddscf1}
\ddot{a}(t)=\e^{-\frac{f_0 (-t+t_s)^{1+\alpha } }{1+\alpha } }f_0 (-t+t_s)^{-1+ \alpha } \left ( f_0(-t+t_s)^{1+\alpha } - \alpha \right )\, .
\end{equation}
Therefore, when $\alpha<1$, the parameter $\ddot{a}$ never vanishes and is always singular, while for $\alpha>1$, it vanishes in two points, namely,
\begin{equation}
\label{rootsddtor}
t=t_s-\left ( \frac{\alpha}{f_0}\right )^{1/(\alpha+1)},{\,}{\,}{\,}t=t_s\, .
\end{equation}
Notice that for $\alpha>1$ the physical system of the scalar field develops a Type IV singularity, as we demonstrated earlier. Let us elaborate on these cases, to see the physical consequences these bring along. For the case $\alpha>1$, if inflation ends at $t=t_s$, the physical system has a Type IV singularity and a singular $w_\mathrm{eff}$ at $t=t_s$, with inflation ending at $t=t_s$, if $t_s$ is considered to occur early enough. A more interesting case occurs if $t_s$ is considered to occur at late times, and inflation ends at a time $t_e$, with $t_e$,
\begin{equation}
\label{te}
t_e=t_s-\left ( \frac{\alpha}{f_0}\right )^{1/(\alpha+1)}\, .
\end{equation}
This can be true if $f_0\gg \alpha$. In this case, inflation ends at $t_e$, which might be quite earlier than $t_s$, and at $t_s$ the Universe experiences a Type IV singularity. Interestingly enough, if inflation ends at the Type IV singularity solution $t=t_s$ given in Eq.~(\ref{rootsddtor}), which occurs if $\alpha>1$, then after $t>t_s$, the effective equation of state describes quintessence acceleration. We have to mention, however, that, if we include quantum effects, the vacuum might acquire higher energy and consequently, inflation might occur again. So the analysis we performed is valid at a classical level, and in order to avoid such issues which originate from the quantum fluctuations of the scalar field, the hybrid inflation scenario has to be used \cite{sergnoj}. For the hybrid inflation consequences, the reader is referred to Ref.~\cite{sergnoj}, where this issue has been studied in detail.

Before closing this section, we shall briefly mention how the presence of perfect matter fluids affects the cosmological singularities of the physical system that consists of a scalar field with Hubble rate given in Eq.~(\ref{IV1}). In this case, the scalar field potential and $\omega(\varphi )$ become \cite{sergnoj},
\begin{align}
\label{ma10-1}
\omega(\phi)=& - \frac{2}{\kappa^2}f'(\phi)-(w_m+1)F_0 \e^{-3(1+w_m)F(\phi)}\, , \nn
V(\phi)=&\frac{1}{\kappa^2}\left(3f(\phi)^2 + f'(\phi)\right)+\frac{w_m-1}{2}F_0 \e^{-3(1+w_m)F(\phi)}\, ,
\end{align}
where $F_0$ an integration constant and $F'(\phi)=f(\phi)$. From Eq.~(\ref{ma10-1}) it easily follows that the exponential in both $V(\phi )$ and $\omega(\phi )$, has a milder singularity structure, in comparison to the rest singular terms, for the Hubble parameter given by Eq.~(\ref{IV1}). Consequently, the same analysis we performed previously applies and in addition notice that the potential is not of a power law type, therefore it deviates from our primary motivation for our study, which is power-law potentials, so we omit this study for brevity. A detailed analysis on this issue was performed in Ref.~\cite{sergnoj}.

\section{Unification of Type IV-singular inflation with de Sitter dark energy epoch leading to a Type IV future singularity}

As we already mentioned in the introduction, it is possible that the Universe may have passed from a Type IV singularity between the singular inflation era and the $\Lambda$CDM epoch. In addition, in the context of unified inflation-dark energy models given in Ref.~\cite{Nojiri:2005pu}, it is possible to have a future singularity of Type II, III or IV, because of the epoch of singular inflation, and in effect we may have a de Sitter-like or pure $\Lambda$CDM dark energy era. In view of these evolutionary aspects, in this section we shall provide an example, in which, we may have a unified description of singular inflation and quintessential dark energy epoch, leading to a Type IV singularity. The model we shall present is particularly fine-tuned to realize this scenario, nevertheless it is an example of a scalar-tensor theory leading to singular inflation and quintessential type of dark energy with a Type IV singularity at finite time. For a recent scenario in the spirit of ours, see \cite{Geng:2015fla}.

We shall assume that the function $f(\phi)$, defined in Eq.~(\ref{ma11}), is equal to,
\begin{equation}\label{ex1hub}
f(t)=c_1(-t+t_f)^{\alpha}+c_2(-t+t_s)^{\beta}+c_1t^{\alpha}+g\, ,
\end{equation}
where $g$ is an arbitrary constant. In addition, we impose the condition that the physical system of the scalar field described by Eq.~(\ref{ex1hub}), experiences a Type IV singularity at the time $t_f$ which is considered to occur at the end, or right after the inflationary era.

In addition, it is also required that the Universe experiences an additional Type IV singularity at the time $t_s$, which is considered to occur much more later than the present time, which we denote $t_p$. So in effect, we have the following conditions,
\begin{equation}\label{cond1}
t_s\gg t_p\gg t_f\, .
\end{equation}
Thus the model (\ref{ex1hub}) describes singular inflation plus a dark energy era, which we now study in detail. Since, the two singularities that occur are of Type IV, the parameters $a$ and $b$ must satisfy the following constraints,
\begin{equation}\label{cond2}
\alpha >1\, , \quad \beta >1\, ,
\end{equation}
and also we shall assume that $a$ and $b$ are integers. Using the reconstruction method of the previous section, the kinetic function $\omega(\phi)$ and the scalar potential $V(\phi )$ of the corresponding scalar-tensor theory are,
\begin{align}\label{potentandomega}
\omega (\phi )=&-\frac{2 \left(-\alpha c_1 (t_f-\phi )^{-1+\alpha}-\beta c_2 (t_s-\phi )^{-1+\beta}+\alpha c_1 \phi ^{-1+\alpha}\right)}{\kappa ^2} \, , \nn
V(\phi )=&\frac{-\alpha c_1 (t_f-\phi )^{-1+\alpha}-\beta c_2 (t_s-\phi )^{-1+\beta}+\alpha c_1 \phi ^{-1+\alpha}+2 \left(g+c_1 (t_f-\phi )^{\alpha}+c_2 (t_s-\phi )^{\beta}+c_1 \phi ^{\alpha}\right)^2}{\kappa ^2}\, ,
\end{align}
while the corresponding EoS defined in Eq.~(\ref{eos}), is equal to,
\begin{equation}\label{eosexample}
w_\mathrm{eff}=-1-\frac{2 \left(\alpha c_1 t^{-1+\alpha}-\alpha c_1 (-t+t_f)^{-1+\alpha}-\beta c_2 (-t+t_s)^{-1+\beta}\right)}{3 \left(g+c_1 t^{\alpha}+c_1 (-t+t_f)^{\alpha}+c_2 (-t+t_s)^{\beta}\right)^2}\, .
\end{equation}
Let us consider the case that $\alpha$ is equal to,
\begin{equation}\label{acase1}
\alpha=2m+1\, ,
\end{equation}
thus it is an odd integer. Near the Type IV singularity at $t\sim t_f$, the EoS (\ref{eosexample}) behaves as follows,
\begin{equation}\label{eosbehavinfsing}
w_\mathrm{eff}=-1-\frac{2 \left(\alpha c_1 t^{-1+\alpha}-\beta c_2 (-t+t_s)^{-1+\beta}\right)}{3 \left(g+c_1 t^{\alpha}+c_2 (-t+t_s)^{\beta}\right)^2}\, ,
\end{equation}
since the following two terms vanish when $t\sim t_f$,
\begin{equation}\label{termsvanish}
c_1 (-t+t_f)^{\alpha}\simeq 0,{\,}{\,}{\,}\alpha c_1 (-t+t_f)^{-1+\alpha}\simeq 0\, ,
\end{equation}
since $\alpha >1$. The expression given in Eq.~(\ref{eosbehavinfsing}) can be further simplified, by considering the fact that $t\ll t_s$, when $t\sim t_f$, and also that $\beta >1$, in conjunction with the fact that $-t+t_s\simeq t_s \gg 1$. So eventually, Eq.~(\ref{eosbehavinfsing}) can be rewritten as follows,
\begin{equation}\label{eosbehavinfsing1}
w_\mathrm{eff}=-1-\frac{2 \left(\alpha c_1 t^{-1+\alpha}-\beta c_2t_s^{-1+\beta }\right)}{3 \left(g+c_1 t^{\alpha}+c_2 t_s^{\beta}\right)^2}\, .
\end{equation}
Notice the term appearing in the numerator of the fraction in Eq.~(\ref{eosbehavinfsing1}). This term is negative, because $t_s\gg t$ and $\beta >1$. Therefore, the EoS of state is $w_\mathrm{eff}>-1$, for $t\sim t_f$, which describes quintessential acceleration. We now proceed to examine the behavior of the EoS for cosmic times corresponding to present time $t_p$. So for $t\sim t_p$, since $t\gg t_f$, and owing to $\alpha$ being equal to an odd integer (\ref{acase1}), the following two terms appearing in Eq.~(\ref{eosexample}) can be approximated as follows,
\begin{equation}\label{approxofterms2}
(-t+t_f)^{\alpha-1}\simeq (-t)^{\alpha-1}=t^{\alpha-1}\, , \quad (-t+t_f)^{\alpha}\simeq (-t)^{\alpha}=-t^{\alpha}\, ,
\end{equation}
where we used the fact that since $a$ is odd integer ($\alpha >1$), $\alpha-1$ is even. Thus, the EoS can be written,
\begin{equation}\label{eosexamplepresenttimeapprox1}
w_\mathrm{eff}=-1+\frac{ 2\beta c_2 (-t+t_s)^{-1+\beta}}{3 \left(g+c_2 (-t+t_s)^{\beta}\right)^2}\, ,
\end{equation}
which describes quintessential acceleration. By taking into account that we assumed $t_s\gg t_p$ and also that $t$ is of the order $t\sim t_p$, the equation above can be further simplified, to the following expression,
 \begin{equation}\label{eosexamplepresenttimeapprox2}
w_\mathrm{eff}=-1+\Lambda_1\, ,
\end{equation}
where we have set,
\begin{equation}\label{lambda1set}
\Lambda_1=\frac{ 2\beta c_2 t_s^{-1+\beta}}{3 \left(g+c_2 t_s^{\beta}\right)^2}=\mathrm{const.}\,
\end{equation}
By choosing $g\gg 1$ and since $t_s\gg 1$, in conjunction to the fact that $\beta >1$, the constant parameter is approximately zero, $\Lambda_1\simeq 0$, and therefore the EoS becomes nearly $w_\mathrm{eff}\simeq -1$, which is de Sitter acceleration. Notice that since we assumed $t\sim t_p$, with $t_p$ corresponding to present time, this era is the dark energy, late time era. So it is possible to end up to a de Sitter dark energy era.

Coming back to Eq.~(\ref{eosexample}), let us study how the EoS behaves near the future Type IV singularity that occurs for cosmic times of the order $t\sim t_s$. Owing to the fact that $t\sim t_s$ and also since $t\gg t_f$, Eq.~(\ref{eosexample}) can be approximated as follows,
\begin{equation}\label{eosexamplefuturesing}
w_\mathrm{eff}=-1-\frac{2 \left(\alpha c_1 t^{-1+\alpha}-\alpha c_1 (-t)^{-1+\alpha}\right)}{3 \left(g+c_1 t^{\alpha}+c_1 (-t+t_f)^{\alpha}\right)^2}\, ,
\end{equation}
since for $t\sim t_s$ the following term vanishes,
\begin{equation}\label{asxeto}
-\beta c_2 (-t+t_s)^{-1+\beta}\simeq 0\, ,
\end{equation}
Owing to the fact that, for an odd $\alpha$, we have $(-t)^{\alpha-1}=t^{\alpha-1}$, the second term in Eq.~(\ref{eosexamplefuturesing}) vanishes, so the EoS is approximately equal to $w_\mathrm{eff}\simeq -1$. Consequently, near the Type IV singularity, the Universe accelerates in a nearly de Sitter way. In principle, the Universe could survive the future Type IV singularity, so we could also study the behavior after the future singularity, but we omit this, for brevity. We only highlight the possibility that the Universe, after the future Type IV singularity can evolve to phantom acceleration, if $\beta$ is equal to some even integer (which would imply that $\beta-1$ is an odd integer). Indeed, the corresponding EoS, which for $t>t_s$ is equal to,
\begin{equation}\label{effeqnstateafterfuture}
w_\mathrm{eff}=-1+\frac{2 \beta c_2 (-t+t_s)^{-1+\beta}}{3 \left(g+c_2 (-t+t_s)^{\beta}\right)^2}\, ,
\end{equation}
which is $w_\mathrm{eff}<-1$, since for $t>t_s$ and with $\beta-1$ being an odd integer, the following holds true,
\begin{equation}\label{dgedf}
2 \beta c_2 (-t+t_s)^{-1+\beta}=-2 \beta c_2 (t-t_s)^{-1+\beta}\, .
\end{equation}
Recall that we assumed $\alpha =2m+1$, and this was the most physically appealing case. The other case, $\alpha =2m$ yields similar results, but there is no possibility to have de Sitter dark energy, so we omit it for brevity, since the analysis is the same as in the case we just studied.

Therefore, in this section we presented a scalar-tensor model, according to which, the Universe experiences the following:
\begin{itemize}

\item A Type IV singular inflation

\item Possible quintessential or approximately de Sitter dark energy era

\item A future Type IV singularity after the dark energy era

\end{itemize}
There are more possibilities to study, in the context of this model, for example if someone assumes that the future Type IV singularity occurs near the present time, that is $t_p\simeq t_s$. Also it would be interesting to find a pure modified gravity description with this type of behavior. We hope to address these issues in a future work.

\section{$F(R)$-gravity analysis}

Having provided a quantitative analysis of the singularity structure in the context of singular inflation, with special emphasis in the Type IV singularity, we now study how the physical system under study can be generated by an $F(R)$ gravity. Particularly, we shall be interested in which $F(R)$ gravity generates the Type IV singularity, that is, how the $F(R)$ gravity behaves near the Type IV singularity. For a similar approach to our reconstruction study, see Ref.~\cite{sergbam08}.  Note that we can equivalently assume that the singularity occurs at the end or after inflation, or even at late times, since our approach is based on the existence of the Type IV singularity and also works as the cosmic time approaches the finite time where the singularity occurs. Therefore, having the Jordan frame $F(R)$ gravity at hand, we can calculate the inflationary parameters to check the viability of the $F(R)$ gravity, but we defer this to a future work. Before we start off, it is worth describing in brief the underlying theoretical framework. For reviews and important papers on the subject, the reader is referred to \cite{reviews,importantpapers}. The general $F(R)$ gravity action, with matter fluids present, is given by,
\begin{equation}
\label{action}
\mathcal{S}=\frac{1}{2\kappa^2}\int \mathrm{d}^4x\sqrt{-g}F(R)+S_m(g_{\mu \nu},\Psi_m)\, ,
\end{equation}
where $\kappa^2=8\pi G$ and in addition, $S_m$ describes the matter fluids action. We shall adopt the metric formalism approach \cite{reviews}, in the context of which, by varying the action with respect to the metric $g_{\mu \nu}$, we obtain the following equations of motion,
\begin{align}
\label{modifiedeinsteineqns}
R_{\mu \nu}-\frac{1}{2}Rg_{\mu \nu}=\frac{\kappa^2}{F'(R)}\left( T_{\mu
\nu}+\frac{1}{\kappa^2} \left(\frac{F(R)-RF'(R)}{2}g_{\mu \nu}+\nabla_{\mu}\nabla_{\nu}F'(R)-g_{\mu
\nu}\square F'(R) \right) \right) \, .
\end{align}
with the prime denoting in this case differentiation with respect to the Ricci scalar $R$ and moreover, $T_{\mu \nu}$ stands for the energy momentum tensor corresponding to the matter fields. In the $F(R)$ modified case, the energy momentum tensor receives an extra contribution originating from the $F(R)$ gravitational sector, with this addition being the feature that renders $F(R)$ gravity a modified version of ordinary Einstein-Hilbert gravity. This additional contribution to the energy momentum tensor is equal to,
\begin{equation}
\label{newenrgymom}
T^\mathrm{eff}_{\mu \nu}=\frac{1}{\kappa}\left( \frac{F(R)-RF'(R)}{2}g_{\mu
\nu}+\nabla_{\mu}\nabla_{\nu}F'(R)-g_{\mu \nu}\square F'(R)\right)\, .
\end{equation}
Finally, in all the following sections, we shall assume a flat FRW metric of the form given in Eq.~(\ref{metricformfrwhjkh}), in which case the Ricci scalar is given by,
\begin{equation}
\label{ricciscal}
R=6(2H^2+\dot{H})\, ,
\end{equation}
Before providing a full quantitative description of the $F(R)$ gravity near the singularity, we shall give a qualitative description of how the $F(R)$ gravity behaves near the singularity. We assume that the $F(R)$ gravity is of the form,
\be
\label{X}
F(R) \sim F_0 + F_1 R^\epsilon\, .
\ee
where $F_0$ and $F_1$ are constants, with $F_0$ an arbitrary constant, allowed to be even zero, but $F_1$ is constrained to be non-zero, that is, $F_1\neq 0$. Then by using the trace part of the $F(R)$ equation,
\be
\label{Scalaron}
3\Box f'(R)= R+2f(R)-Rf'(R)\, .
\ee
we obtain the following relation,
\be
\label{XII}
3 F_1 \Box R^{\epsilon -1} = \left\{
\begin{array}{ll} R & \ \mbox{when $\epsilon<0$ or $\epsilon=2$} \\
\left(2-\epsilon\right) F_1 R^\epsilon & \ \mbox{when $\epsilon>1$ or
$\epsilon\neq 2$}
\end{array} \right. \, .
\ee
Then in the FRW background (\ref{metricformfrwhjkh}), the Hubble rate $H$ behaves as follows,
\be
\label{XIII}
H \sim \frac{h_0}{\left(t_s - t\right)^\beta}\, ,
\ee
where $h_0$ and $\beta$ are constants. Consequently, the scalar curvature
$R=6\dot H + 12 H^2$ behaves as follows,
\be
\label{XIV}
R \sim \left\{ \begin{array}{ll}
\frac{12h_0^2}{\left(t_s - t\right)^{2\beta}} & \ \mbox{when $\beta>1$} \\
\frac{6 h_0 + 12 h_0^2}{\left(t_s - t\right)^2} & \ \mbox{when $\beta=1$} \\
\frac{6\beta h_0}{\left(t_s - t\right)^{\beta + 1}} & \ \mbox{when
$\beta<1$}
\end{array} \right. \, .
\ee
According to Eqs.~(\ref{XIII}) and (\ref{XIV}), it easily follows that the $\beta\geq 1$ case corresponds to the
Type I (Big Rip) singularity in \cite{Caldwell:2003vq,ref5}, and the
$1>\beta>0$ case corresponds to Type III singularity. Moreover the $0>\beta>-1$ case corresponds to Type II while when $\beta<-1$
(but $\beta\neq\mbox{integer}$) we have a Type IV singularity.

We now proceed to the detailed description of the $F(R)$ gravity near the Type IV singularity of Eq.~(\ref{IV1}), by using the reconstruction techniques that were firstly developed in \cite{Nojiri:2006gh,Capozziello:2006dj,Nojiri:2006be}. What we are mainly interested in, is to find which $F(R)$ gravity exactly generates the Hubble rate (\ref{IV1}), for $f_1(t)=0$ and $f_2(t)=f_0$, as $t$ tends to $t_s$. Notice that as $t\rightarrow t_s$, then $t-t_s\rightarrow 0$. The action of a general $F(R)$ gravity, in the absence of matter fluids, is equal to,
\begin{equation}
\label{action1dse}
\mathcal{S}=\frac{1}{2\kappa^2}\int \mathrm{d}^4x\sqrt{-g}F(R)\, .
\end{equation}
By varying the above action with respect to the metric, we obtain the first FRW metric which is equal to,
\begin{equation}
\label{frwf1}
 -18\left ( 4H(t)^2\dot{H}(t)+H(t)\ddot{H}(t)\right )F''(R)+3\left (H^2(t)+\dot{H}(t) \right )F'(R)-\frac{F(R)}{2}=0\, .
\end{equation}
Using an auxiliary scalar field $\phi $, the $F(R)$ gravity action given in Eq.~(\ref{action1dse}) can be written in the following way,
\begin{equation}
\label{neweqn123}
S=\int \mathrm{d}^4x\sqrt{-g}\left ( P(\phi )R+Q(\phi ) \right )\, .
\end{equation}
The main goal of the reconstruction technique is to find the exact form of the functions $P(\phi )$ and $Q(\phi )$, which depend explicitly on the auxiliary scalar field. Particularly, if these functions are expressed as implicit functions of the Ricci scalar, the $F(R)$ gravity will follow easily. In order to find these, we shall use the fact that the auxiliary scalar field $\phi $ is a dynamical (meaning that it depends on time) degree of freedom, auxiliary because it has not kinetic term. Then, by varying the auxiliary field modified action of Eq.~(\ref{neweqn123}) with respect to this auxiliary degree of freedom, we get the following algebraic equation,
\begin{equation}
\label{auxiliaryeqns}
P'(\phi )R+Q'(\phi )=0\, ,
\end{equation}
with the prime this time denoting differentiation with respect to $\phi$. If this equation can be solved explicitly with respect to the auxiliary field $\phi $, will yield the function $\phi (R)$. Correspondingly, the $F(R)$ gravity can easily be obtained by making use of $\phi (R)$ and substituting this to the auxiliary $F(R)$ action of Eq.~(\ref{neweqn123}), and we finally get,
\begin{equation}
\label{r1}
F(\phi( R))= P (\phi (R))R+Q (\phi (R))\, .
\end{equation}
Hence, what we need to do is to find the functions $P(\phi )$ and $Q(\phi )$, given the Hubble rate,
\begin{equation}
\label{hurnew}
H(t)=f_0\left( -t+t_s \right)^{\alpha}\, .
\end{equation}
In order to find the functions $P(\phi )$ and $Q(\phi )$, we vary the action (\ref{neweqn123}) with respect to the metric tensor, and by using a flat FRW metric, we obtain the following differential equation,
\begin{align}
\label{r2}
0= & -6H^2P(\phi (t))-Q(\phi (t) )-6H\frac{\mathrm{d}P\left (\phi (t)\right )}{\mathrm{d}t}=0\, , \nn
0=& \left ( 4\dot{H}+6H^2 \right ) P(\phi (t))+Q(\phi (t) )+2\frac{\mathrm{d}^2P(\phi (t))}{\mathrm {d}t^2}+\frac{\mathrm{d}P(\phi (t))}{\mathrm{d}t}=0\, .
\end{align}
By eliminating $Q(\phi (t))$ from Eq.~(\ref{r2}) we obtain,
\begin{equation}
\label{r3}
2\frac{\mathrm{d}^2P(\phi (t))}{\mathrm {d}t^2}-2H(t)\frac{\mathrm{d}P(\phi (t))}{\mathrm{d}t}+4\dot{H}P(\phi (t))=0\, .
\end{equation}
Therefore, by solving Eq.~(\ref{r3}) with respect to $P(t)$, we obtain the explicit form of $P(t)$, given the Hubble rate. Accordingly, we easily find the function $Q(t)$, by using the first equation of Eq.~(\ref{r2}). Notice the interplay of the variables $\phi$ and $t$, as they appear in the previous equations. As is explicitly proven in  Ref.~\cite{Nojiri:2006gh}, owing to the mathematical equivalence of the $F(R)$ action (\ref{action1dse}) and the auxiliary $F(R)$ action (\ref{neweqn123}), the scalar field is identical to the cosmic time $t$, that is $\phi =t$.

Using this technique, substituting the Hubble rate (\ref{hurnew}) in Eq.~(\ref{r3}), we get the following differential equation,
\begin{equation}
\label{ptdiffeqn}
2\frac{\mathrm{d}^2P(t)}{\mathrm {d}t^2}-2 f_0 (-t+t_s)^{\alpha }\frac{\mathrm{d}P(t)}{\mathrm{d}t}-4 f_0 (-t+t_s)^{-1+\alpha } \alpha P(t)=0\, .
\end{equation}
By setting $x=-t+t_s$, Eq.~(\ref{ptdiffeqn}) can be written in the following way,
\begin{equation}
\label{dgfere}
\frac{\mathrm{d}^2P(x)}{\mathrm {d}x^2}+f_0 x^{\alpha }\frac{\mathrm{d}P(x)}{\mathrm{d}x}-2 f_0 x^{-1+\alpha } \alpha P(x)=0\, .
\end{equation}
This differential equation can be easily solved analytically by substituting $z=x^{\alpha+1}$, in which case the differential equation becomes,
\begin{equation}
\label{diffeqn}
(\alpha+1)^2z\frac{\mathrm{d}^2P(z)}{\mathrm {d}z^2}+(\alpha+1)(f_0 z+\alpha)\frac{\mathrm{d}P(z)}{\mathrm{d}z}-2 f_0 \alpha P(z)=0\, ,
\end{equation}
with general solution,
\begin{equation}
\label{genrealsol}
P(z)= \e^{\frac{\ln(z)-z f_0}{1+\alpha }} C_1 U\left(-\frac{-1-3 \alpha }{1+\alpha },1+\frac{1}{1+\alpha },\frac{z f_0}{1+\alpha }\right) + \e^{\frac{\ln(z)-z f_0}{1+\alpha }} C_2 L_{n}^{m}\left(\frac{z f_0}{1+\alpha }\right)\, ,
\end{equation}
where the parameters $n$ and $m$ are equal to,
\begin{equation}
\label{gfgrg}
n=\frac{-1-3 \alpha }{1+\alpha },{\,}{\,}{\,}m=\frac{1}{1+\alpha }\, ,
\end{equation}
also $C_1,C_2$ are arbitrary constants and finally $U(a,b,z)$ and $L_n^m(z)$ are the confluent Hypergeometric function and the generalized Laguerre polynomial respectively. By substituting again $z=x^{\alpha+1}$ we get,
\begin{equation}
\label{genrealsolxfunct}
P(x)= \e^{\frac{\ln(x^{\alpha+1})-x^{\alpha+1} f_0}{1+\alpha }} C_1 U\left(-\frac{-1-3 \alpha }{1+\alpha },1+\frac{1}{1+\alpha },\frac{x^{\alpha+1} f_0}{1+\alpha }\right) + \e^{\frac{\ln(x^{\alpha+1})-x^{\alpha+1} f_0}{1+\alpha }} C_2 L_{n}^{m}\left(\frac{x^{\alpha+1} f_0}{1+\alpha }\right)\, .
\end{equation}
Having at hand $P(x)$, we can easily compute $Q(x)$, which reads,
\begin{align}
\label{qtanalyticform}
Q(x) =& -\frac{1}{1+\alpha } 6 \e^{-\frac{x^{1+\alpha } f_0}{1+\alpha }} f_0 x^{-1+\alpha }
\left( x^{1+\alpha } \right)^{\frac{1}{1+\alpha }} \left(\left( 1+f_0 x^{1+\alpha }\right) \right. \nn
& \times \left(1+\alpha \right) \left( C_1 U\left( \frac{1+3 \alpha }{1+\alpha },1+\frac{1}{1+\alpha },\frac{x^{1+\alpha } f_0}{1+\alpha }\right) +C_2 L_{n_1}^{m_1}\left( \frac{x^{1+\alpha } f_0}{1+\alpha } \right) \right) -x^{1+\alpha }
\\ \notag & \times \left( \left(1+\alpha \right) C_1 U\left(\frac{1+3 \alpha }{1+\alpha },1+\frac{1}{1+\alpha },\frac{x^{1+\alpha } f_0}{1+\alpha }\right) \right.
\\ \notag & + \left( 1+3 \alpha \right) C_1 U \left( \frac{2+4 \alpha }{1+\alpha },2+\frac{1}{1+\alpha },\frac{x^{1+\alpha } f_0}{1+\alpha }\right) \nn
& \left. \left. + \left( 1+\alpha \right) C_2\left( L_{n_1}^{m_1}\left( \frac{x^{1+\alpha } f_0}{1+\alpha }\right)+L_{n_2}^{m_2}\left( \frac{x^{1+\alpha } f_0}{1+\alpha }\right)\right) \right) f_0 \right)\, ,
\end{align}
with $n_1,m_1$ being equal to,
\begin{equation}
\label{n1m1}
n_1=-\frac{1+3 \alpha }{1+\alpha}\, ,\quad m_1=\frac{1}{1+\alpha }\, ,
\end{equation}
while $n_2$ and $m_2$ are equal to,
\begin{equation}
\label{n2ma}
n_2=-\frac{2+4 \alpha }{1+\alpha }\, , \quad m_2=1+\frac{1}{1+\alpha }\, .
\end{equation}
Substituting $P(x)$ and $Q(x)$ from Eqs.~(\ref{genrealsolxfunct}) and (\ref{qtanalyticform}) in Eq.~(\ref{auxiliaryeqns}) and solving with respect to $x$, one can in principle derive the final form of the function $x(R)$. This task, however, is rather formidable, owing to the complex structure of the functions $P(x)$ and $Q(x)$. In order to obtain an approximate form of the resulting $F(R)$ gravity, we shall exploit the fact that we are interested to know the $F(R)$ gravity near the Type IV singularity, so this means that $x\rightarrow 0$. We therefore expand $P'(x)$ and $Q'(x)$ in small powers of $x$, and by keeping dominant terms (recall that $\alpha>1$ in order a Type IV singularity occurs), the resulting expansion of $P(x)$ reads,
\begin{equation}
\label{pseries}
P(x)\simeq \frac{\mathcal{A}}{x^{\alpha+1}}+\mathcal{O}(x)\, ,
\end{equation}
with the parameter $\mathcal{A}$ being equal to,
\begin{equation}
\label{parametalpha}
\mathcal{A}=\frac{C_1 \Gamma\left(\frac{1}{1+\alpha }\right) \left(\frac{f_0}{1+\alpha }\right){}^{-\frac{1}{1+\alpha }}}{(1+\alpha ) \Gamma\left(\frac{1+3 \alpha }{1+\alpha }\right)}-\frac{(1+3 \alpha ) C_1\Gamma\left(1+\frac{1}{1+\alpha }\right) \left(\frac{f_0}{1+\alpha }\right){}^{-\frac{1}{1+\alpha }}}{(1+\alpha ) \Gamma\left(1+\frac{1+3 \alpha }{1+\alpha }\right)}\, .
\end{equation}
Moreover, the function $Q(x)$ for small values of the argument $x$ can be approximated as,
\begin{equation}\label{qsapprox}
Q(x)\simeq \frac{\mathcal{B}}{x^{\alpha+1}}+\mathcal{C}+\mathcal{O}(x)\, ,
\end{equation}
with the parameters $\mathcal{B}$ and $\mathcal{C}$ being equal to,
\begin{align}
\label{mathbandc}
\mathcal{B}=& \frac{6 f_0 (1+3 \alpha ) C_1 \left((2+\alpha ) \Gamma\left(\frac{3+5 \alpha }{1+\alpha }\right) \Gamma\left(1+\frac{1}{1+\alpha }\right)-2 (1+2 \alpha ) \Gamma\left(\frac{2+4 \alpha }{1+\alpha }\right) \Gamma\left(2+\frac{1}{1+\alpha }\right)\right) \left(\frac{f_0}{1+\alpha }\right){}^{-\frac{1}{1+\alpha }}}{ (1+\alpha )^2 \Gamma\left(\frac{2+4 \alpha }{1+\alpha }\right) \Gamma\left(\frac{3+5 \alpha }{1+\alpha }\right)} \, , \nn
\mathcal{C}=& \frac{6 f_0^{\frac{2\alpha+1}{\alpha+1}} C_1 \left(\frac{(1+\alpha ) (2+\alpha ) \Gamma\left(\frac{1}{1+\alpha }\right)}{\Gamma\left(\frac{1+3 \alpha }{1+\alpha }\right)}-\frac{\alpha  (1+3 \alpha )^2 \Gamma\left(1+\frac{1}{1+\alpha }\right)}{\Gamma\left(\frac{2+4 \alpha }{1+\alpha }\right)}-\frac{4 \left(1+5 \alpha +6 \alpha ^2\right) \Gamma\left(1+\frac{1}{1+\alpha }\right)}{\Gamma\left(\frac{2+4 \alpha }{1+\alpha }\right)}+\frac{4 (1+2 \alpha )^2 (1+3 \alpha ) \Gamma\left(2+\frac{1}{1+\alpha }\right)}{(2+\alpha ) \Gamma\left(\frac{3+5 \alpha }{1+\alpha }\right)}\right) }{(1+\alpha )^{3+\alpha}}\, .
\end{align}
Substituting Eqs.~(\ref{pseries}) and (\ref{qsapprox}) in Eq.~(\ref{auxiliaryeqns}), we finally get,
\begin{equation}
\label{finalxr}
x\simeq \left(-\frac{\mathcal{C}}{\mathcal{A}R+\mathcal{B}}\right)^{\frac{1}{\alpha+1}}\, .
\end{equation}
Finally, by substituting (\ref{finalxr}) in Eq.~(\ref{r1}), we obtain the final form of the $F(R)$ gravity that generates the Type IV singularity, which is,
\begin{equation}
\label{finalfrgravity}
F(R)\simeq -\frac{\mathcal{A}^2}{\mathcal{C}}R^2-2\frac{\mathcal{B}\mathcal{A}}{\mathcal{C}}R-\frac{\mathcal{B}^2}{\mathcal{C}}+\mathcal{C}\, .
\end{equation}
Notice that this $F(R)$ gravity can be brought to a form of Einstein-Hilbert gravity plus curvature corrections, if the free parameter $C_1$ is chosen appropriately. Indeed, if,
\begin{equation}
\label{cond1-B}
-2\frac{\mathcal{B}\mathcal{A}}{\mathcal{C}}=1\, ,
\end{equation}
which can be true if $C_1$ is equal to,
\begin{align}\label{c1valueforeinsteihilber}
& C_1=-\frac{\mathcal{C}(1+\alpha )^2 \Gamma\left(\frac{2+4 \alpha }{1+\alpha }\right) \Gamma\left(\frac{3+5 \alpha }{1+\alpha }\right)}{2\mathcal{A}6 f_0 (1+3 \alpha ) \left((2+\alpha ) \Gamma\left(\frac{3+5 \alpha }{1+\alpha }\right) \Gamma\left(1+\frac{1}{1+\alpha }\right)-2 (1+2 \alpha ) \Gamma\left(\frac{2+4 \alpha }{1+\alpha }\right) \Gamma\left(2+\frac{1}{1+\alpha }\right)\right) \left(\frac{f_0}{1+\alpha }\right){}^{-\frac{1}{1+\alpha }}}\, ,
\end{align}
then, the $F(R)$ gravity is,
\begin{equation}\label{einsteinlikegrav}
F(R)\simeq R-\frac{\mathcal{A}^2}{\mathcal{C}}R^2-\frac{\mathcal{B}^2}{\mathcal{C}}+\mathcal{C}\, .
\end{equation}
Thus, the resulting $F(R)$ gravity near the Type IV singularity can be brought to an Einstein-Hilbert form, plus curvature corrections and cosmological constant. Notice that the final form of the $F(R)$ gravity near the singularity is of the form (\ref{X}), that is, of polynomial type.

Actually, the form of $F(R)$ gravity is a nearly $R^2$ gravity plus cosmological constant, when $\frac{\mathcal{A}^2}{\mathcal{C}}<0$. In order to see this, we change the notation of the coefficients in Eq.~(\ref{einsteinlikegrav}) as follows,
\begin{equation}\label{coeffcnew}
C_0=-\frac{\mathcal{C}}{4\mathcal{A}^2},{\,}{\,}{\,}\Lambda =-\frac{\mathcal{B}^2}{\mathcal{C}}+\mathcal{C}\, .
\end{equation}
By doing so, the Jordan frame $F(R)$ gravity takes the following form,
\begin{equation}\label{jordanframegravity}
F(R)=R+\frac{R^2}{4C_0}+\Lambda \, .
\end{equation}
This Jordan frame theory is equivalent to a variant of the Starobinsky model \cite{starobinsky} in the Einstein frame. The pure Jordan frame $F(R)$ gravity is of the form,
\begin{equation}
\label{pure}
\mathcal{S}=\frac{1}{2\kappa^2}\int \mathrm{d}^4x\sqrt{-\hat{g}}F(R)\, ,
\end{equation}
Starting from action (\ref{pure}), and by introducing the auxiliary field $A$, the action (\ref{pure}) can be cast in the following form,
\begin{equation}\label{action1dse111}
\mathcal{S}=\frac{1}{2\kappa^2}\int \mathrm{d}^4x\sqrt{-\hat{g}}\left ( F'(A)(R-A)+F(A) \right )\, .
\end{equation}
By varying (\ref{action1dse111}) with respect to the auxiliary field $A$, we easily obtain the solution $A=R$, and consequently this verifies the mathematical equivalence of the actions (\ref{pure}) and (\ref{action1dse111}). We can be transferred to the Einstein frame by making the canonical transformation,
\begin{equation}\label{can}
\varphi =-\sqrt{\frac{3}{2\kappa^2}}\ln (F'(A)) \, ,
\end{equation}
where $\varphi$ is the canonical Einstein frame scalar field (or inflaton field). By conformally transforming the Jordan frame metric,
\begin{equation}\label{conftransmetr}
g_{\mu \nu}=e^{-\varphi }\hat{g}_{\mu \nu } \, ,
\end{equation}
with the ``hat'' denoting the Jordan frame metric, we get the Einstein frame canonical scalar field action,
\begin{align}\label{einsteinframeaction}
\mathcal{\tilde{S}}= & \int \mathrm{d}^4x\sqrt{-g}\left ( \frac{R}{2\kappa^2}-\frac{1}{2}\left (\frac{F''(A)}{F'(A)}\right )^2g^{\mu \nu }\partial_{\mu }A\partial_{\nu }A -\frac{1}{2k^2}\left ( \frac{A}{F'(A)}-\frac{F(A)}{F'(A)^2}\right ) \right ) \nn
=& \int \mathrm{d}^4x\sqrt{-g}\left ( \frac{R}{2k^2}-\frac{1}{2}g^{\mu \nu }\partial_{\mu }\varphi\partial_{\nu }\varphi -V(\varphi )\right )\, .
\end{align}
The scalar potential $V(\varphi )$, which is a function of the canonical scalar field $\varphi $ is equal to,
\begin{align}\label{potentialvsigma}
V(\varphi )=\frac{A}{F'(A)}-\frac{F(A)}{F'(A)^2}=\frac{1}{2\kappa^2}\left ( e^{\sqrt{2\kappa^2/3}\varphi }R\left (e^{-\sqrt{2\kappa^2/3}\varphi} \right )- e^{2\sqrt{2\kappa^2/3}\varphi }F\left [ R\left (e^{-\sqrt{2\kappa^2/3}\varphi} \right ) \right ]\right ) \, .
\end{align}
Upon using the form of the Jordan frame $F(R)$ gravity (\ref{jordanframegravity}), the scalar potential of Eq.~(\ref{potentialvsigma}) becomes \cite{sergeistarobinsky},
\begin{equation}\label{vapprox}
V(\varphi)\simeq C_0+C_2e^{-2\sqrt{\frac{2}{3}}\kappa   \varphi }+C_1e^{-\sqrt{\frac{2}{3}}\kappa   \varphi } \, ,
\end{equation}
with the coefficients $C_1$ and $C_2$ standing for,
\begin{equation}\label{coeffnewdefs}
C_1=-2C_0,{\,}{\,}{\,}C_2=C_0-\Lambda \, .
\end{equation}
The scalar potential (\ref{vapprox}) describes a nearly Starobinsky Einstein frame canonical scalar theory. If the coefficient $C_2$ is equal to $C_0$, then the potential (\ref{vapprox}) becomes exactly the Einstein frame Starobinsky model, namely,
\begin{equation}\label{einsteinfrstarbky}
V(\varphi )=C_0\left (1-e^{\sqrt{\frac{2}{3}}\kappa \varphi}\right )^2\, .
\end{equation}

\section{Impact of Type IV and finite time singularities on scalar perturbative modes}

In this section we shall briefly discuss the impact of finite time singularities on the power spectrum of perturbed modes,
with special emphasis to the Type IV singularities. A thorough analysis on this issue however was presented in \cite{barrowsing}, with
regards to the classical stability of sudden singularities. In addition, in \cite{barrowsing} the Big Rip case was also extensively discussed.
Notice that the Type IV singularity falls in the category of some sudden singularities and before we proceed to a qualitative analysis of the model
we studied in this paper, it is worth outlining the most important results of \cite{barrowsing}. For details on this important issue, the reader
is referred to \cite{barrowsing} and related references therein. As was shown in \cite{barrowsing}, if $a(t)\sim (t_s-t)^{\lambda}$, the
energy density and pressure are equal to $\rho \sim (t_s-t)^{2\lambda -2}$ and $p\sim (t_s-t)^{\lambda -2}$. Then if $\lambda >1$ and if $\lambda $ is
a non-integer number, all the scalar metric perturbations are bounded. Let us make some correspondence of the finite time singularities with sudden
singularities. The case $1<\lambda <2$ describes the Type II singularity, so the results of \cite{barrowsing} indicate that singularities like Type
II or milder can provide us with bounded scalar metric perturbations. Actually, the Type IV singularity is milder than the Type II, so it is
expected that the scalar metric perturbations are bounded in the Type IV case too. In the rest of this section we shall perform a brief
qualitative analysis for the case at hand, with special emphasis to the long-wavelength modes and for a Type IV singularity.

\subsection{Long-wavelength solutions during and after inflation: The Type IV case}

As is very well known \cite{inflation,mukh1,perturbations}, the quantum fields which are present during the primordial acceleration of the Universe, experience quantum mechanical fluctuations. These primordial fluctuations could serve as a very crucial observational feature and a test of any potentially viable cosmological theory. This is owing to the fact that the primordial perturbations could have a direct impact on the present epoch Universe's large scale structure. In principle, the kinetic energy of the scalar fields is responsible for the generation of a density distribution at a given spacelike hypersurface, during the primordial acceleration. The perturbations of this density distribution are responsible, to a great extent, for the large scale structure in the Universe, materialized by gravitational instabilities.

Consider a homogeneous, isotropic and spatially flat Universe. The quantity that the determines the scale that within it causal physical processes can be materialized, is the sound Hubble radius $c_sH^{-1}$, where $H$ is the Hubble rate. The parameter $c_s$ stands for the sound speed, which for a canonical scalar field $c_s=1$ (this corresponds to the case we study in this paper). The sound Hubble radius plays a crucial role towards the understanding of the way that the spectrum of primordial perturbations is actually generated. Another physically important quantity is the fraction $k/aH$, with $a$ the scale factor and $k$ the wavenumber. As the primordial acceleration occurs, the scale factor increases rapidly, and consequently the comoving Hubble radius $(Ha)^{-1}$, decreases. During the same epoch, the perturbation's physical length , satisfies the relation $a/k \ll c_sH^{-1}$. The latter relation is very important since this actually determines the way that the scalar modes are evolving. We shall quantify this evolution later in this section, but let us qualitatively analyze what is the meaning of this. Practically, the relation $a/k \ll c_sH^{-1}$ indicates that the physical scales grow more quickly in comparison to the Hubble sound radius and in effect, the scalar modes we study, begin their evolution within the horizon $c_sH^{-1}$. In the process of the accelerating expansion, the perturbation modes (scalar field quantum fluctuations), become larger more quickly than the sound Hubble radius. Consequently at some point it passes out of the horizon, so it freezes out. This freezing has a direct impact on the scale dependence of the scalar perturbations, at the time this freezing occurs, which is when the Hubble sound horizon is crossed. Practically, the modes at that point enter the long-wavelength limit $aH\gg c_sk$. In the context of this approximation, what plays a crucial role for the determination of the scale dependence, is the equation of state of the matter fluids present at moment of the horizon crossing, which can occur for example during the matter or during the post recombination era. In the following we are not interested in determining the evolution of the mode per se, for a specific matter fluid present, but our aim is to study the effect of a Type IV singularity on the evolution of a long-wavelength frozen mode, with emphasis to the question whether the modes are bounded or not.

Assume that the singularity appears at the end of inflation, so $t_s$ indicates the time that inflation ends. The scale factor corresponding to the Hubble rate of Eq.~(\ref{IV1}), with $f_1(t)=0$ and $f_2(t)=f_0$, namely $H(t)=f_0 \left( t_s - t \right)^\alpha $, is equal to,
 \begin{equation}\label{hublaweqns}
a(t)=e^{-\frac{f_0 (-t+t_s)^{1+\alpha }}{1+\alpha }} \, .
\end{equation}
What is the main focus in this section is to study the behavior of the scalar perturbations during and after the inflationary era and qualitatively check whether the existence of a Type IV singularity at the end of the inflationary era affects the spectrum of primordial scalar perturbations. The perturbed metric that will yield the scalar perturbations, is of the following form,
\begin{equation}\label{metricperturbations}
\mathrm{d}s^2=\left( 1+2\phi \right)\mathrm{d}t^2-a^2(1-2\psi )\sum_i\mathrm{d}x_i^2 \, ,
\end{equation}
where $\phi$ and $\psi$ quantify the metric perturbation. Notice that in order to obtain the scalar field perturbation, we also perturb the canonical scalar field $\varphi $, in the following way, $\varphi (t,\vec{x})'=\bar{\varphi}+\delta \varphi (t,\vec{x})$. Note that $\vec{x}$ denotes all the space coordinates and $\bar{\varphi}$ denotes a homogeneous scalar field. Since at first order in perturbation theory, $\phi=\psi$, in order to obtain the metric perturbations only, we are left practically with two variables, the variable $\phi$ and the inflaton variation $\delta \varphi$, which are related through the Einstein equations. It can be shown \cite{mukh1} that
the equation which describes the evolution of one scalar mode $\phi_k$, with wavenumber $k$, takes the following form,
\begin{equation}\label{basicequation}
\ddot{\phi_k}+\left(H-2\frac{\ddot{\bar{\varphi}}}{\dot{\bar{\varphi}}}\right)\dot{\phi_k}-\left(2H\frac{\ddot{\bar{\varphi}}}{\dot{\bar{\varphi}}}
+8\pi G\dot{\bar{\varphi}}^2-\frac{k^2}{a(t)^2}\right)\phi_k=0 \, ,
\end{equation}
with $a (t)$ being the scale factor. We shall be interested in the long-wavelength solutions of this equation, which corresponds to the case $aH\gg c_sk$. Since the speed of sound for a canonical scalar field is equal to one, the limiting case at hand corresponds to wavelengths with $aH\gg k$. It is possible to show that during any stage, the long-wavelength ($k\rightarrow 0$) analytic solution of Eq.~(\ref{basicequation}) takes the following form,
\begin{equation}\label{solution}
\phi_k=C_1(k)\left (1-\frac{H}{a}\int ^ta(t)\mathrm{d}t\right )-C_2(k)\frac{4\pi G}{k^2}\frac{H}{a} \, ,
\end{equation}
with $C_1(k)$ the coefficient of the non-decaying modes, while $C_2(k)$, the coefficient of the decaying modes. Using the scale factor of Eq.~(\ref{hublaweqns}), we easily obtain the following form of the long-wavelength mode $\phi_k$,
\begin{align}\label{longwave}
\phi_k= & -\frac{4 C_2(k) e^{\frac{f_0 (-t+t_s)^{1+\alpha }}{1+\alpha }} f_0 G \pi  (-t+t_s)^{\alpha }}{c k^2}
\nn
 &+C_1(k) \left(1-\frac{e^{\frac{f_0 (-t+t_s)^{1+\alpha }}{1+\alpha }} f_0 (-t+t_s)^{1+\alpha } \left(\frac{f_0 (-t+t_s)^{1+\alpha }}{1+\alpha }\right)^{-\frac{1}{1+\alpha }} \Gamma\left[\frac{1}{1+\alpha },\frac{f_0 (-t+t_s)^{1+\alpha }}{1+\alpha }\right]}{1+\alpha }\right) \, .
\end{align}
It is obvious that the existence of a Type IV singularity does not cause any inconsistencies to the long-wavelength modes, with inconsistencies meaning any type of singular or abnormal behavior. Therefore we may conclude that the Type IV singularities are harmless. A qualitatively similar conclusion was derived in \cite{Barrow:2015ora}  (see also \cite{barrowsing}), with the corresponding transition to the standard cosmological model not affecting the power spectrum.

\section{Brief discussion of finite time singularities and the graceful exit of inflation}

One of the most serious problems that the first models of inflation \cite{guth} were confronted with, was the exit from the inflationary era, a problem known as the graceful exit. Initially, in the context of the false vacuum theories of inflation, it was not possible to consistently describe a mechanism for the smooth transition of an inflationary Universe to the classical FRW expansion. This problem was solved in the new inflation theories \cite{linde,steinhard}, where the slow-roll approximation actually solved the problem of graceful exit. Indeed, if the inflaton scalar slowly rolls down the potential, towards the minimum of the potential, then inflation ends where the slow-roll approximation breaks down, and the field oscillates around the minimum of it's potential. In the case we studied in this paper, the canonical scalar field potential of
Eq.~(\ref{varphipotential}) belongs to a class of potentials were slow-roll applies. Therefore, the graceful exit issues in our case is dealt within the context of power-law inflation models, which are in good agreement with observations \cite{planck}.

However, the exit of inflation in our case becomes a serious problem when the slow-roll approximation breaks down. As we demonstrated in a previous section, this is true when the Type IV singularity occurs during the inflationary era, since the slow-roll parameters become divergent. Also the $e$-folding number becomes divergent at the time the singularity occurs. Therefore, the problematic case is only when the singularity occurs during the inflationary era. If however the singularity occurs at the end or much later than the inflationary era, the slow-roll approximation is not violated and therefore the graceful exit can be dealt in the context of the power law models of inflation. If the singularity occurs much more later than the ending of inflation (even during the dark energy era), then slow-roll breaks when the inflationary indices become of order one, and the exit from inflation goes in the standard way, for example, the resulting de Sitter Universe becomes unstable and the exit is guaranteed.

Before we close this brief discussion, it is worth to note another quite physical appealing but exotic new possibility that a singularity during inflation might generate. As we demonstrated in the text, in the power-law models of inflation of the form (\ref{varphipotential}). when the singularity occurs during the inflationary era, the slow-roll approximation is violated in a violent way, since the slow-roll parameters become divergent and also the $e$-folding number it self becomes divergent. If however the violation of the slow-roll approximation is not divergent, this could be viewed as another mechanism for a graceful exit from inflation. So the appearance of a singularity during inflation could generate instability in the inflationary spacetime, and this could indicate a new mechanism for a consistent exit from inflation era. This is another view to the Type IV singularities and work is in progress towards this direction.

\section*{Conclusions}

We have thoroughly examined the existence of classical finite time singularities in the context of scalar-tensor theories. Using a very generic behavior for the Hubble rate, we provided a quantitative description of all the finite time singularities of the physical system under study, with emphasis to the Type IV singularity. Particularly, with regards to the Type IV singularity, we investigated when this can occur and we also showed what is the impact of this singularity on the slow-roll inflation parameters. As we explicitly demonstrated, the presence of this type of singularity may have dramatic effects on the slow-roll parameters, rendering them singular, in the case that the singularity occurs during the inflationary era. After exemplifying our claims with illustrative examples, we studied the behavior of the effective equation of state parameter and in which cases, its value implies acceleration. In addition, we thoroughly examined all the cases at which the Universe might experience a finite time singularity, with the singularity being during, at the end, after inflation. In addition, we studied the case at which the singularity occurs much more later than inflation ends. The issue of when inflation classically ends was examined too. As we showed, in the case that a Type IV singularity is present, the results are of some interest, especially in the particular case that the time $t_s$, at which the singularity occurs, is much more later in comparison to the time $t_e$, at which inflation ends.

In the context of the above scenarios, the Universe experiences a Type IV singularity at finite time $t_s$. Interestingly enough, the Type IV singularity is not a crushing singularity, such as the Big Rip singularity is for example. This means that at that point, extensibility and geodesic incompleteness do not necessarily occur. With regards to this, the nature of the Type IV singularity needs yet to be understood at a fundamental level. Usually in nature, non extensible singularities are protected, as in the case of spacetime singularities in black holes which are dressed with the horizons of black holes. So having at large scales undressed, naked, crushing singularities, is a rather inconvenient feature of classical cosmological theories. Perhaps the crushing singularities occur as a result of our classical approach and maybe it is the signal that a quantum gravitational description is needed to amend these issues in a concrete way. In fact, in the context of loop quantum gravity, initial crushing singularities do not occur, see for example \cite{LQC}. However, the Type IV singularity is not a severe spacetime singularity, so it is of critical importance to find a consistent explanation for the occurrence of unbounded values of the higher derivatives of the Hubble rate, an also critically examine if this feature can be considered harmful for the theory itself. A first step towards this direction was attempted in this paper, where we showed that if the Type IV singularity occurs during inflation, this can be catastrophic for the slow-roll parameters and the corresponding observational indices. But if this Type IV singularity occurs at the end, or right after the end of inflation, compatibility with the observational data can be achieved to some extent.

Another feature that has to be addressed, in the context of finite time singularities, is the issue of including quantum corrections, in the spirit of Ref.~\cite{Nojiri:2005sx}. For the case of the Type IV singularity, quantum effects would be negligible, owing to the fact that the curvature is not growing significantly there. However, the quantum effect of conformal matter around the Type I, II and Type III singularities should be taken into account. As was shown in \cite{Nojiri:2005sx}, the singularity structure gets milder in the presence of quantum corrections, so the same is expected in the present case too.

Finally a comment is in order. Having the possibility of a singular inflation, makes in some way the need for more accurate observational data more compelling. This is due to the fact that in this way, the various choices of a consistent theoretical description of inflation will eventually be narrowed down. With the present paper, we have excluded scalar-tensor models which cause a Type IV singularity during inflation, since these would be in conflict with current observational data, and also we have set the stage for new investigations towards this line of research. Therefore, in the absence of an ultimate theory of quantum gravity that will explain all the singularity issues, even a better understanding of non-crushing singularities, like the Type IV one, is of critical importance.

\section*{Acknowledgments}

This work is  supported  in part by the JSPS Grant-in-Aid for Scientific
Research (S) \# 22224003 and (C) \# 23540296 (S.N.) and in part by MINECO (Spain), projects FIS2010-15640 and FIS2013-44881.

\end{document}